\documentclass[%
 reprint,
 amsmath,amssymb,
 aps,
longbibliography
]{revtex4-1}

\usepackage{graphicx}
\usepackage{dcolumn}
\usepackage{bm}
\usepackage{tikz}
\usepackage[mathlines]{lineno}
\usepackage[utf8]{inputenc}
\usepackage{xcolor}

\usepackage{scalerel}
\usepackage{tikz}
\usetikzlibrary{svg.path}

\definecolor{orcidlogocol}{HTML}{A6CE39}
\tikzset{
  orcidlogo/.pic={
    \fill[orcidlogocol] svg{M256,128c0,70.7-57.3,128-128,128C57.3,256,0,198.7,0,128C0,57.3,57.3,0,128,0C198.7,0,256,57.3,256,128z};
    \fill[white] svg{M86.3,186.2H70.9V79.1h15.4v48.4V186.2z}
                 svg{M108.9,79.1h41.6c39.6,0,57,28.3,57,53.6c0,27.5-21.5,53.6-56.8,53.6h-41.8V79.1z M124.3,172.4h24.5c34.9,0,42.9-26.5,42.9-39.7c0-21.5-13.7-39.7-43.7-39.7h-23.7V172.4z}
                 svg{M88.7,56.8c0,5.5-4.5,10.1-10.1,10.1c-5.6,0-10.1-4.6-10.1-10.1c0-5.6,4.5-10.1,10.1-10.1C84.2,46.7,88.7,51.3,88.7,56.8z};
  }
}

\newcommand\orcidicon[1]{\href{https://orcid.org/#1}{\mbox{\scalerel*{
\begin{tikzpicture}[yscale=-1,transform shape]
\pic{orcidlogo};
\end{tikzpicture}
}{|}}}}

\usepackage[colorlinks=true,allcolors=blue]{hyperref}

\makeatletter
\renewcommand*\env@matrix[1][\arraystretch]{%
  \edef\arraystretch{#1}%
  \hskip -\arraycolsep
  \let\@ifnextchar\new@ifnextchar
  \array{*\c@MaxMatrixCols c}}
\makeatother

\begin{document}

\preprint{APS/123-QED}

\preprint{APS/123-QED}

\title{Optimization of Flat to Round Transformers with Self-fields using Adjoint Techniques}

\author{L. Dovlatyan}
 \email{levondov@umd.edu}
\author{B.L. Beaudoin\orcidicon{0000-0001-9935-4658}, 
S. Bernal\orcidicon{0000-0001-8287-6601}, 
I. Haber, 
D. Sutter, and 
T.M. Antonsen Jr.\orcidicon{0000-0002-2362-2430}}

\collaboration{Institute For Research in Electronics and Applied Physics, University of Maryland, College Park, Maryland 20742, USA}

\date{\today}

\begin{abstract}
A continuous system of moment equations is introduced that models the transverse dynamics of a beam of charged particles as it passes through an arbitrary lattice of quadrupoles and solenoids in the presence of self-fields. Then, figures of merit are introduced specifying system characteristics to be optimized. The resulting model is used to optimize the parameters of the lattice elements of a flat to round transformer with self-fields, as could be applied in electron cooling. Results are shown for a case of no self-fields and two cases with self-fields. The optimization is based on a gradient descent algorithm in which the gradient is calculated using adjoint methods that prove to be very computationally efficient. Two figures of merit are studied and compared: one emphasizing radial force balance in the solenoid, the other emphasizing minimization of transverse beam energy in the solenoid.
\end{abstract}

\maketitle


\section{INTRODUCTION}

Beams of charged particles in accelerators are guided and manipulated by complex systems of magnets whose design optimization requires tracking the trajectories of the particles through the fields of the magnets as well as through the self-fields of the beam. Since the properties of the beam after passing through a lattice of magnetic focusing elements depend in a complicated way on the many parameters describing the lattice and self-fields, optimizing the lattice is a computationally intensive task, and historically, a large body of knowledge and techniques for addressing the optimization problem have been developed.

Traditionally, the design of these magnet systems is carried out using computer codes that calculate the beam particle phase space trajectories in the prescribed lattice of magnets.  The process often reduces to the optimization of figures of merit (FoMs) in the high dimensional parameter space characterizing the lattice.  Because of the large number of parameters, the efficiency of the optimization algorithm is critical.  One class of optimization algorithms, based on calculating the gradient of the FoM in parameter space, becomes computationally prohibitive if the gradient is to be calculated directly (by individually varying each of the parameters) in the high dimensional parameter space. 
    
The number of computations needed to calculate the gradient can be reduced via the introduction of adjoint techniques \cite{Director1969,nikolova2006,jameson1995,antonsen2019,antonsen2019_2}.  In this approach an alternate, but related, mathematical problem is introduced in which in a single (or in several) computation(s) the linear dependence of the FoM on all the parameters can be determined.  The adjoint approach has previously been applied in circuit theory \cite{Director1969}, electromagnetics \cite{nikolova2006}, aerodynamics \cite{jameson1995}, and accelerator physics \cite{antonsen2019,antonsen2019_2}, as well as in other fields. 
    
\begin{figure}[b]
\includegraphics[width=0.45\textwidth]{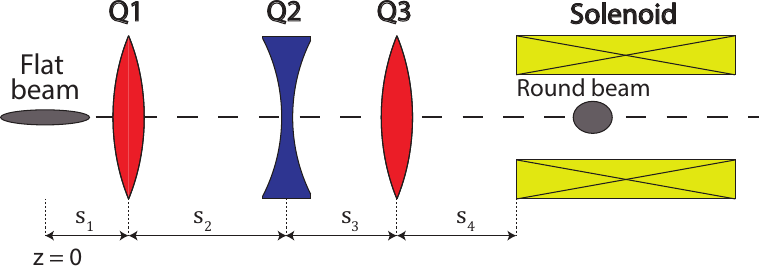}
\caption{General geometry of quadrupole triplet for FTR beam transformations.}
\label{fig:1}
\end{figure}

In this paper we will illustrate the application of the adjoint approach to the design of Flat-to-Round (FTR) or Round-to-Flat (RTF) transformers as have been proposed for use in relativistic electron cooling \cite{Derbenev1998,Burov1998,Burov1998_2,Burov2002}.  As the names suggest these transformers are systems of magnets that will convert an unmagnetized beam that has a high aspect ratio, elliptical spatial cross section, to a round beam in a solenoidal magnetic field, or vice versa.  In its simplest form this conversion is accomplished with a triplet of quadrupoles, and a solenoid.  See Fig. \ref{fig:1}.  Parameters that can be varied to optimize this conversion are the positions and strengths of the four magnet elements, including the orientations of the quadrupoles (11 parameters including the location of the first quadrupole).  The method we will introduce also allows for more detailed optimization in terms of the spatial profiles of the magnet elements.  In addition, one might have the option to vary the parameters of the incoming beam, but we will take these to be given.  
The description of the beam that we will use is based on the solution of second moment equations describing the four-dimensional transverse phase space of the beam.  There are 10 such moments corresponding to the ten independent elements of the matrix of second moments, the so-called sigma matrix \cite{Chao1999}.  The effect of self-fields is included in the calculations under the assumption that the beam maintains an elliptical spatial cross section of uniform density.  A future publication will extend the approach to a kinetic or phase space description.  The FoM will be a weighted sum of combinations of the moments at the exit of the transformer. Different combination can be used depending on the desired application.  We will find that combining the adjoint calculation with a gradient descent algorithm allows us to effectively optimize these different FoMs. 

We chose FTR and RTF transformers as an example because of their multiple applications. The original motivation for the development of circular mode adapters, introduced by Derbenev in 1998 \cite{Derbenev1998}, was to improve the rates and efficiency of relativistic electron cooling \cite{Derbenev1998,Burov1998,Burov1998_2,Burov2002} of hadron beams.  Here a high energy electron beam and a hadron beam are made to co-propagate during which the beams interact and the hadrons are cooled.  High energy electron beam sources are generally electron storage devices, in which the beams tend to become flat through radiation effects and intra-beam scattering.  The cooling of the hadron beam is optimized when the electron beam interacting with the hadron beam is magnetized and matches the hadron beam size and shape, which is approximately round, and has a low effective transverse temperature \cite{Derbenev2000}. The adapters provide the means for converting an unmagnetized beam, to the “magnetized” state in the cooling solenoid and then converting back to the unmagnetized state at the exit of the solenoid for reinjection into the storage device.  

Almost immediately after Derbenev's initial proposal, it was realized that an RTF adapter following an electron source immersed in a solenoid field could be used to produce a flat, uncoupled beam of high enough emittance ratio and charge to potentially replace the high energy electron damping rings required in proposed designs for high energy linear colliders \cite{Brinkmann1999}. Experiments at Fermilab and DESY confirmed the feasibility \cite{Brinkmann2001,Edwards2000,Edwards2001}.  Since then, the possible use of circular mode adapters for phase space manipulations have been expanded to include generating round beams at select points in circular colliders to compensate for beam-beam effects at high energy and to offset the effects of space-charge and the cyclotron component of the ion beam in low energy cooler rings \cite{Burov2002}.  Additionally, their use has been explored for injection into high intensity X-ray Sources \cite{Wang2003,Lidia2003} or into dielectric laser driven accelerators \cite{Ody2016,Cropp2019}.  

The FTR and RTF beam transformations typically involve a solenoid and three, 45-degree skew quadrupoles, as depicted in Fig. \ref{fig:1}. In the absence of self-fields, the parameters of the solenoid and the quadrupoles can be determined using transfer matrices. The overall length of the triplet, excluding the solenoid, scales inversely with the strength of the solenoidal field and linearly with the momentum of the beam, as reflected in the reference length
\begin{equation} \label{eq:1}
\beta_s = \frac{2p}{|qB|}
\end{equation}
where $p$ is the momentum, $q$ the charge, and $B$ the maximum solenoidal field. In symmetrical triplets, the simplest configuration, the two outer quadrupoles have the same strength and polarity and are equidistant from the central one with opposite polarity. In asymmetric triplets \cite{Burov1998_2,Burov2002}, the two outer quadrupoles have differing strengths and different separations from the inner quadrupole. 

Using a thin lens model for the symmetric triplet, the distance $d$ between the outer thin quadrupoles $Q_1$, $Q_3$ and the central one $Q_2$, and the inverse focal lengths, $q_1 = q_3, q_2$, are, following Wolske, all given in terms of $\beta_s$ \cite{Wolski2006}: 
\begin{equation} \label{eq:2}
d = \frac{\beta_s}{2\sqrt{1+\sqrt{2}}} \, , \, q_1 = q_3 = \frac{-\sqrt{2}+1}{\beta_s} \, , \, q_2 = \frac{2\sqrt{2}}{\beta_s}.
\end{equation}
In this design, the flat beam (for FTR transformer) is incident directly on the first quadrupole which is located at $z = 0$, i.e., no drift space is assumed between the input beam and the first quadrupole and between the last quadrupole and the solenoid ($s_1=s_4=0$ in Fig \ref{fig:1}). More general equations, ascribed to Edwards \cite{Thrane2002,Sun2005}, still apply to thin lenses and no initial/final drift spaces, but relax the condition of symmetry, i.e., $q_1 \neq q_3 \, , \, s_2 \neq s_3$.  For the triplet described by Eq. (\ref{eq:2}) an initial horizontal flat beam is transformed into a round beam with a canonical angular momentum of opposite sign to the canonical angular momentum term generated by the solenoid on transport to its center.

As an example, Burov and Danilov \cite{Burov1998} described a symmetrical FTR triplet for an electron energy of 500 MeV and solenoid field strength $B_s = 10.3$ kG; the triplet length is 2.23 m.  Any initial triplet thin lens design calculation can use Eq. (\ref{eq:2}), following a choice of the $\beta_s$ parameter in Eq. (\ref{eq:1}). Alternatively, it is possible to scale existing designs, as in references \cite{Burov1998} or \cite{Burov1998_2}, to a desired electron energy and solenoid field. We will adapt as a starting point for the test model to be used in the following scaled version of the Burov-Danilov symmetrical FTR triplet \cite{Burov1998}, which includes non-zero drift spaces $s_1 = s_4$ (Fig. \ref{fig:1}), an electron energy of 5 keV and a solenoid field of 15 G ($\beta_s = 0.319$ m). The resulting overall length of the triplet is 0.213 m. From these initial parameters it is straight forward to obtain a thick lens model that provides the initial conditions for the adjoint optimizations.

The initial FTR optical parameters also provide a computation of the initial beam second moments. If $M_{\textrm{st}}$  is the matrix that transports an initial flat beam through the rotated triplet to the center of the solenoid, then for an initial matrix, $\sum_0$, the sigma matrix at the center of the solenoid is 
\begin{equation} \label{eq:2-1}
\sum = M_{\textrm{st}} \textstyle \sum_0 M_{\textrm{st}}^T
\end{equation}
where $M_{\textrm{st}}^T$ is the transpose \cite{Chao1999,kim2003}. The generalized second moment equations derived below for the use in the adjoint optimization replace the elements of $\sum$ in Eq. \ref{eq:2-1}.

The treatments discussed so far assume no space-charge or self-field effects.  Conservation laws have been developed that relate initial and final beam parameters that are independent of space charge \cite{kim2003}.  However, the design of a system of quadrupoles needed to transform a flat beam to a cylindrically symmetric beam does depend on space charge, and is addressed here. The starting point of the optimizations presented below are the flat to round parameters obtained in the absence of self-fields.

In Sec. II we will introduce a general formulation based on a set of (10) differential equations for the evolution with distance of the second moments of the beam particle distribution function.  The system allows for the profiles of the magnetic fields of the focusing elements to be treated; it allows for the orientations, locations and the strength of the quadrupoles to be varied; and it treats self-fields in the approximation that the beam maintains an elliptical transverse cross section of constant density.  We will also introduce a system of adjoint equations that allows for the efficient calculation of the change in system performance with changing parameters.  In Sec. III we will use the adjoint calculation of the gradient in parameter space to optimize the parameters of a selected model of an FTR transformer under a variety of assumptions and constraints. Here we will also address the role of constants of motion in determining final values for beam parameters.

\section{basic model}

In this section we first present a system of equations that describes the evolution with distance of the second moments of a charged particle beam distribution in the presence of a combination of transverse forces. These forces include the Lorentz force of a spatially varying solenoidal (axial) magnetic field, the Lorentz force of a superposition of arbitrarily oriented quadrupole magnetic fields, and the electric and magnetic self-force due to the beam's charge and current densities.  These moment equations will be used to simulate the propagation of a beam through a system of magnets, which converts a beam with an elliptical cross section to one with a round cross section.  A general figure of merit will be introduced that quantifies how successfully the shape conversion has been made.  Subsequently, we will formally perturb this system by making small changes in the parameters defining the focusing forces and assess the changes by evaluating the change in the figure of merit.  Such an evaluation would be used in a gradient-based optimization scheme.  Then we will introduce an adjoint system of equations that will allow one to calculate compactly the changes in the system due to changes in the focusing parameters.

We note that the basic system of moment equations is well studied, and codes that solve for the moments exist in the community \cite{rangarajan1989}. However, our goal is to linearize the system with respect to magnet parameters including the resulting changes to the self-fields, introduce an adjoint set of equations that efficiently calculate gradients of figures of merit, and perform optimizations.  This requires presenting the base equations, their linearization, and the adjoint system for completeness.  Accordingly, we have relegated much of the detail to appendices.

The moments we consider are averages of products of all possible pairs of variables describing the transverse displacement of beam particles and the rate of change of the transverse displacement with distance.  These moments correspond to the 16 elements of the 4 by 4 transverse sigma matrix mentioned in Eq. (\ref{eq:2-1}).  Due to the symmetry of this matrix only 10 elements are independent.  Thus, our governing system consists of 10 moment evolution equations.  Although there are 10 separate moment equations, we show that as expected there are accompanying conservation laws relating the moments.  The evolution with distance of the sigma matrix is normally treated by matrix multiplication with individual matrices representing focusing elements and drift spaces.  We choose to deal with differential equations for the continuous moments, as this allows us to introduce adjoint equations that include self-field effects and spatial profiles of focusing fields.

The underlying assumption that will be made in deriving the moment equations is that the beam particles' trajectories are well described by the paraxial equations of motion in which transverse forces are linear in the particles' displacements from the axis, or are linearly proportional to a particle's transverse velocity.  The final equations will describe the beam's evolution in the Larmor, or rotating, frame defined by the applied solenoidal magnetic field.  We begin the calculation in the lab frame, transform to the Larmor frame, and complete the derivation in the Larmor frame.  Beam quantities in the lab frame can always be recovered from the Larmor frame values by application of rotation transformations.  These are given in Appendix \ref{A.2}.

To start, we write equations for the evolution of the transverse particle displacements in a Cartesian coordinate system in the lab frame,
\begin{subequations}
\begin{eqnarray}
x_1'' = k_x + k_\Omega y_1' + \frac{1}{2}k_\Omega' y_1 , \label{eq:3a}
\\
y_1'' = k_y - k_\Omega x_1' - \frac{1}{2}k_\Omega'x_1 \label{eq:3b}.
\end{eqnarray}
\end{subequations}
Here ($x_1,y_1$) are a particle's transverse displacements in the lab frame, for example, $x_1$ is the horizontal displacement from the center-line and $y_1$ is the vertical displacement.  A prime denotes differentiation with respect to the axial coordinate $z$, which measures distance along the center-line.  The quantity
\begin{equation}
\label{eq:4}
k_\Omega(z) = \frac{qB_z(z)}{mc\gamma\nu_z}
\end{equation}
is the spatial gyration rate due to the axial (solenoidal) magnetic field. The quantities in Eq. (\ref{eq:4}) are as follows: $q$ and $m$ are the particle charge and mass, $B_z$ is the axial component of the solenoidal magnetic field strength, $c$ is the speed of light, $\nu_z$ is the axial velocity and $\gamma=(1-\nu_z^2/c^2)^{-1/2}$.  (If the beam consists of electrons, $q = -e$, and if $B_z > 0$ then $k_\Omega < 0$.)  The quantities $k_x,k_y$ represent the transverse forces from the quadrupoles and space-charge, and will be defined subsequently.  The second terms in Eq. (\ref{eq:3a}) and (\ref{eq:3b}) describe the Lorenz force due to the axial component of the solenoidal field and the transverse velocity, and the third terms describe the Lorentz force due to the radial component of the solenoidal field and the axial velocity.

Our next step is to transform the variables to their Cartesian representation in a rotating frame,
\begin{equation} \label{eq:5}
\begin{pmatrix} x_1 \\ y_1 \end{pmatrix} = \begin{bmatrix} \cos{\phi} & -\sin{\phi} \\ \sin{\phi} & \cos{\phi} \end{bmatrix} \begin{pmatrix} x \\ y \end{pmatrix}
\end{equation}
where $\phi(z)$ is an axially dependent rotation phase that we choose to satisfy
\begin{equation} \label{eq:6}
\phi' = -\frac{k_\Omega}{2}
\end{equation}
and which defines the Larmor frame.  Substituting Eqs. (\ref{eq:5}) and (\ref{eq:6}) into (\ref{eq:3a}),(\ref{eq:3b}) and multiplying by the inverse of the rotation matrix appearing in Eq. (\ref{eq:5}) results in the system
\begin{equation} \label{eq:7}
\begin{pmatrix} x'' \\ y'' \end{pmatrix} + \left( \frac{k_\Omega}{2} \right)^2 \begin{pmatrix} x \\ y \end{pmatrix} = \begin{bmatrix} \cos{\phi} & \sin{\phi} \\ -\sin{\phi} & \cos{\phi} \end{bmatrix} \begin{pmatrix} k_x \\ k_y \end{pmatrix}.
\end{equation}

We wish to obtain equations for the average of products of the four variables ($x,x',y,y'$).  There are 16 ordered products.  However, order does not matter, leaving 10 independent products.  We choose combinations of these products that distinguish $x$-$y$ symmetric and non-symmetric motion.  For spatial moments we choose
\begin{eqnarray} \label{eq:8}
\bm{Q} = \begin{pmatrix}[1.5] Q_+ \\ Q_- \\ Q_x \end{pmatrix} = \begin{pmatrix}[1.5] \langle x^2 + y^2 \rangle / 2 \\ \langle x^2 - y^2 \rangle / 2 \\ \langle xy \rangle \end{pmatrix}.
\end{eqnarray}
Here the angle brackets imply average over the beam distribution function.  Accompanying these spatial moments are momentum like moments
\begin{equation} \label{eq:9}
\bm{P} = \frac{d}{dz}\bm{Q} = \begin{pmatrix}[1.5] P_+ \\ P_- \\ P_x \end{pmatrix} = \begin{pmatrix}[1.5] \langle xx' + yy' \rangle \\ \langle xx' - yy' \rangle \\ \langle yx' + xy' \rangle \end{pmatrix} .
\end{equation}
The angular momentum completes this group of momentum like moments,
\begin{equation} \label{eq:10}
L = \langle xy' - yx' \rangle.
\end{equation}
Note, that $L$ is the angular momentum in the rotating frame.  The group of 10 moments is completed by three energy-like moments
\begin{eqnarray} \label{eq:11}
\bm{E} = \begin{pmatrix}[1.5] E_+ \\ E_- \\ E_x \end{pmatrix} = \begin{pmatrix}[1.5] \langle x'^2 + y'^2 \rangle  \\ \langle x'^2 - y'^2 \rangle  \\ 2 \langle y'x' \rangle \end{pmatrix}. \\ \nonumber
\end{eqnarray}
The evolution of these moments is determined by the following system of equations
\begin{subequations}
\label{eq:12}
\begin{eqnarray}
\frac{d}{dz}\bm{Q} &=& \bm{P} \, , \label{eq:12a}
\\
\frac{d}{dz}\bm{P} &=& \bm{E} + \bm{O}\cdot\bm{Q} \, , \label{eq:12b}
\\
\frac{d}{dz}\bm{E} &=& \bm{O}\cdot\bm{P} + \bm{N}L \, , \label{eq:12c}
\\
\frac{d}{dz}L &=& -\bm{N}^{\dagger}\cdot\bm{Q} \label{eq:12d}
\end{eqnarray}
\end{subequations}
Here the matrix $\bm{O}$ and vector $\bm{N}$ are defined as follows
\begin{widetext}
\begin{equation} \label{eq:13}
\bm{O} = -\frac{k_\Omega^2}{2} \begin{bmatrix}[1.5] 1&0&0\\0&1&0\\0&0&1 \end{bmatrix} + 2 \sum_{\textrm{quads}}K_q \begin{bmatrix}[1.5] 0&c_q&-s_q\\c_q&0&0\\-s_q&0&0 \end{bmatrix} + \frac{\Lambda}{Q_\Delta} \begin{bmatrix}[1.5] 1&c_\alpha&s_\alpha\\c_\alpha&1&0\\s_\alpha&0&1 \end{bmatrix}
\end{equation}
\end{widetext}
and
\begin{equation} \label{eq:14}
\bm{N} = 2\sum_{\textrm{quads}}K_q \begin{pmatrix}[1.5] 0\\s_q\\c_q \end{pmatrix} - \frac{\Lambda}{Q_\Delta} \begin{pmatrix}[1.5] 0\\s_\alpha\\-c_\alpha \end{pmatrix}.
\end{equation}

The matrices defined in Eqs. (\ref{eq:13}) and (\ref{eq:14}) are continuous functions of axial distance, $z$. The first matrix in Eq. (\ref{eq:13}) is due to the solenoidal field, the second is due to the quadrupoles, and the third matrix is due to the self-fields.  The first vector in Eq. (\ref{eq:14}) is due to the quadrupoles, and the second is due to the self-fields.  In Eqs. (\ref{eq:13}) and (\ref{eq:14}) the following expressions and notation have been introduced.  Each quadrupole magnet has field strength given in the lab frame by
\begin{eqnarray} \label{eq:15}
\nonumber
B_{qx} &=& B_q'(z) \left[ \sin{(2\psi_q)}x_1 - \cos{(2\psi_q)}y_1 \right] \\[0.1cm]
B_{qy} &=& -B_q'(z) \left[ \cos{(2\psi_q)}x_1 + \sin{(2\psi_q)}y_1 \right]
\end{eqnarray}
where $B_q'(z)$ defines the strength and axial profile of the quadrupole field and the angle $\psi_q$ defines the orientation. As shown in Appendix \ref{A.1}, this leads to the displayed contributions to the $\bm{O}$ matrix and $\bm{N}$ vector where
\begin{equation} \label{eq:16}
K_q(z) = \frac{qB_q'(z)}{mc\gamma \nu_z}.
\end{equation}
If $K_q > 0$ the magnet will be defocusing in the lab frame when $\theta = \psi_q,\psi_q+\pi$, and focusing when $\theta=\psi_q+\pi/2,\psi_q-\pi/2$. Here, $\theta$ is the angle a particle's transverse displacement makes with respect to the $x_1$ axis. If $K_q < 0$ these are reversed. In the Larmor frame the orientation of the quadrupole is characterized by the variables
\begin{equation} \label{eq:17}
(s_q,c_q) = \left(\,\sin{(2\phi-2\psi_q)},\cos{(2\phi-2\psi_q)}\,\right)
\end{equation}
where $\phi(z)$ is the angle resulting from the integration of Eq. (\ref{eq:4}).  We note that with one group of quadrupoles placed separately from the solenoid it is always possible to define the phase such that it vanishes in the region of the quadrupoles.

The self-field term is calculated in the Larmor frame in Appendix \ref{A.2}.  Here it is assumed that the beam charge density distribution is spatially uniform inside an ellipse with differing major and minor radii and tilted at an angle in the Larmor frame.  The values of the major and minor radii and the angle are determined by the three spatial moments ($Q_+,Q_-,Q_x$).  The result is the following
\begin{subequations}
\label{eq:18}
\begin{eqnarray}
Q_\Delta &=& \left[Q_+^2-(Q_-^2+Q_x^2)\right]^{1/2} \, , \label{eq:18a}
\\
c_\alpha &=& - \frac{Q_-}{(Q_++Q_\Delta)} \, , \label{eq:18b}
\\
s_\alpha &=& - \frac{Q_x}{(Q_++Q_\Delta)} \label{eq:18c} .
\end{eqnarray}
\end{subequations}
The strength of the self-fields is measured by the beam current parameter
\begin{equation} \label{eq:19}
\Lambda = \frac{cqZ_0I}{4\pi m \nu_z^3 \gamma_0^3}
\end{equation}
with $Z_0=377$ Ohms, and $I$ is the beam current.

The equations (\ref{eq:12a}) - (\ref{eq:12d}) can be combined to show the following conservation relation
\begin{equation} \label{eq:20}
\frac{d}{dz} \left[ \bm{E} \cdot \bm{Q} + \frac{L^2}{2} - \frac{1}{2} \bm{P} \cdot \bm{P} \right] = 0.
\end{equation}
This does not represent the conservation of a four-dimensional emittance, $\epsilon_{4D} = \textrm{det}[\Sigma]$. Rather, it is the conserved quantity $I_2(\Sigma)$ given by Eq. (12) of \cite{rangarajan1989} and attributed to \cite{Lebedev_2010}. The implications of these constants will be discussed subsequently.

\begin{figure}[t]
\includegraphics[width=0.45\textwidth]{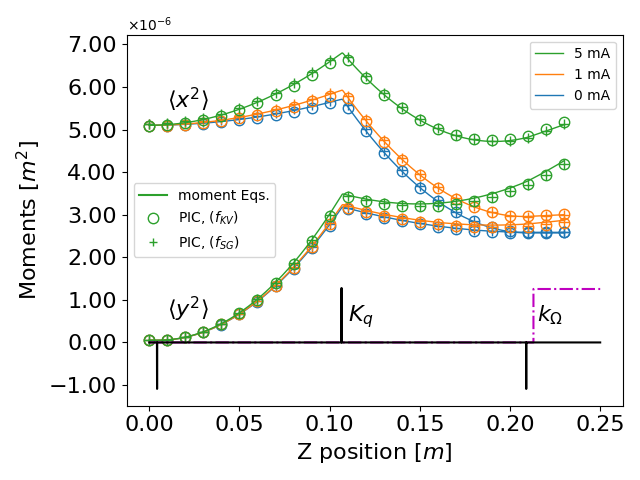}
\caption{Comparison of solutions of the moment  equations (solid lines) with solutions from the PIC code Warp (circular and cross symbols). PIC simulations were run with an initial K-V beam distribution (circles) \cite{kv1959} as well as a Semi-Gaussian (crosses) distribution using 40K particles. Comparisons are for beams with 0, 1, and 5 mA current.}
\label{fig:2}
\end{figure}

We compare the solution of the moment equations with the moments calculated by the PIC code Warp \cite{friedman2013} in Fig. \ref{fig:2}.  For this simulation we consider a symmetric triplet with parameters selected to convert a flat beam to a round one in the presence of space-charge.  The space-charge is evaluated along the entire path of the beam.  Parameters for the simulation are given in Table \ref{tab:1}.  The large pipe radius in the list of parameters is chosen to minimize the contributions from image charge forces on the simulations in order to compare the results with the moment equations.

\begin{table}[b]
\caption{PIC code simulation and FTR lattice parameters.}
\begin{ruledtabular}
\begin{tabular}{lc}
\textrm{Parameter}&
\textrm{Value} \\
\colrule
Beam distribution & K-V, Semi-Gaussian \\
Particle number & 40000 \\
Longitudinal step-size & 0.1 [mm] \\
Grid cell number & 2048 x 2048 \\
Pipe radius & 250.0 [mm] \\
\colrule
Beam energy & 5 keV \\
Quadrupole 1 location & 0.0043 [m] \\
Quadrupole 2 location & 0.1066 [m] \\
Quadrupole 3 location & 0.2090 [m] \\
Quadrupole 1,3 strength & -18.236 [T/m] \\
Quadrupole 2 strength & 21.364 [T/m] \\
Quadrupole 1,2,3 length & 0.1 [mm] \\
Solenoid location & 0.2133 \\
Solenoid strength & $15 \times 10^{-4}$ [T] \\
Initial $\sqrt{\langle x^2 \rangle}, \sqrt{\langle y^2 \rangle}$ & 2.2581, 0.2258 [mm] \\
Initial rms $\epsilon_x,\epsilon_y$\footnote{where $\epsilon_x = \sqrt{2\langle x^2 \rangle \langle x'^2 \rangle - \langle xx' \rangle ^2}$, $\epsilon_y = \sqrt{2\langle y^2 \rangle \langle y'^2 \rangle - \langle yy' \rangle ^2}$ \label{f1}} & 16,0.16 [mm-mrad] 
\end{tabular}
\end{ruledtabular}
\label{tab:1}
\end{table}
Plotted in Fig. \ref{fig:2} are the computed values of $x_{\textrm{rms}}^2=\langle x^2\rangle=Q_++Q_-$ and $y_{\textrm{rms}}^2=\langle y^2\rangle=Q_+-Q_-$ as functions of $z$ for three values of beam current.  The solutions of the moment equations are solid lines, and the PIC values are shown as circles and crosses. For the PIC case two different distributions are used: a K-V \cite{kv1959} distribution function:
\begin{equation} \label{eq:20-2}
    f_{KV}(x,x',y,y') = \frac{1}{4\pi^2A}\delta \Big(\frac{x^2}{\langle x^2 \rangle} + \frac{y^2}{\langle y^2 \rangle} + \frac{x'^2}{\langle x'^2 \rangle} + \frac{y'^2}{\langle y'^2 \rangle} - 4\Big)
\end{equation} \\*[0.01cm]
where $A = \sqrt{\langle x^2 \rangle \langle y^2 \rangle \langle x'^2 \rangle \langle y'^2 \rangle}$ and $\delta$ is the Dirac delta function, as well as a Semi-Gaussian distribution function:
\begin{subequations} \label{eq:20-3}
    \begin{align}
    f_{SG}(x,x',y,y') = \frac{1}{8\pi^2A} \exp{\Big( -\frac{1}{2}(\frac{x'^2}{\langle x'^2 \rangle} + \frac{y'^2}{\langle y'^2 \rangle}) \Big)}
    \end{align}
\end{subequations}
when $x^2 / \langle x^2 \rangle + y^2 / \langle y^2 \rangle < 4$ otherwise $f_{SG} = 0$. Where a Semi-Gaussian is uniform in space over the cross section of the beam and Gaussian in velocity, and represents a distribution function from a thermionic cathode. As can be seen in Fig. \ref{fig:2}, in the zero current case the beam becomes round at the end of the triplet, whereas in the presence of self-fields the roundness is spoiled.  In the next section we will show how the optimizations of the triplet parameters can recover the round state of the beam.

We now imagine that Eqs. (\ref{eq:12a}) - (\ref{eq:12d}) have been solved for a given set of initial conditions, $(\bm{Q},\bm{P},\bm{E},L)\big|_{z_i}$, and magnetic field parameters and profiles, which we label with a vector $\bm{a}$. An assessment of the configuration can be made on the basis of a figure of merit that depends on the final values of the moments,
\begin{equation} \label{eq:21}
F(\bm{Q},\bm{P},\bm{E},L, \bm{a})\Big|_{z_f} \, .
\end{equation}
For a flat to round transition, the figure of merit might be a function which when minimized forces the components $Q_x$, $Q_-$ and $\bm{P}$ all to zero, signifying the beam is round and its radius will remain constant. Alternatively, the FoM might force the beam to be cylindrically symmetric in the solenoid with no rotation or minimum second moment $E_{+,\textrm{Lab}} = \langle \, x_1'^2 \,+\, y_1'^2 \,\rangle$. This will be discussed in more detail in the next section.  We refer to a general solution just described (not necessarily the minimizing one) as a base case.

If a small change is made in the focusing parameters, $\delta\bm{a}$, there will be a commensurately small change in the moments as functions of z, and a small change in the figure of merit from the corresponding base case.  The small changes in the moments will satisfy a linearization of Eqs. (\ref{eq:12}) - (\ref{eq:14}).  We denote the small changes with a superscript ($X$), and we refer to them as true changes.  These satisfy
\begin{subequations}
\label{eq:22}
\begin{align}
\frac{d}{dz}\delta \bm{Q}^{(X)} &= \delta \bm{P}^{(X)} \label{eq:22a}
\\*[0.1cm]
\frac{d}{dz}\delta \bm{P}^{(X)} &= \delta \bm{E}^{(X)} + \bm{O} \cdot \delta \bm{Q}^{(X)} + \delta \bm{O}^{(X)} \cdot \bm{Q} \label{eq:22b}
\\*[0.1cm]
\frac{d}{dz} \delta \bm{E}^{(X)} &= \bm{O} \cdot \delta \bm{P}^{(X)} + \bm{N} \delta L^{(X)} + \delta \bm{O}^{(X)} \cdot \bm{P} + \delta \bm{N}^{(X)}L \label{eq:22c}
\\*[0.1cm]
\frac{d}{dz} \delta L^{(X)} &= -\bm{N}^{\dagger} \cdot \delta \bm{Q}^{(X)} - \delta \bm{N}^{\dagger(X)} \cdot \bm{Q} \label{eq:22d}
\end{align}
\end{subequations}
The resulting change in the figure of merit is 
\begin{align} \label{eq:23}
\delta F^{(X)} = \bigg[ \delta \bm{Q}^{(X)} \frac{\partial F}{\partial \bm{Q}} + \delta \bm{P}^{(X)} \frac{\partial F}{\partial \bm{P}} 
\nonumber \\*[0.1cm] + \delta \bm{E}^{(X)} \frac{\partial F}{\partial \bm{E}} + \delta L^{(X)} \frac{\partial F}{\partial L} \bigg]_{z_f} + \delta \bm{a} \frac{\partial F}{\partial \bm{a}} \bigg|_{Q,P,E,L}.
\end{align}
Again, the superscript $(X)$ signifies that it is a true perturbation, i.e. the result of changing the focusing system: the solenoidal and quadrupole magnetic fields.

The change in the FoM due to changes in parameters has two types of contribution: contributions due to the explicit dependence of the FoM on parameters, the last term on the right of Eq. (\ref{eq:23}), and contributions due to the implicit dependence on parameters due to changes in the moments.

The perturbed matrices $\bm{O}^{(X)},\bm{N}^{(X)}$ appearing in Eqs. (\ref{eq:22b}) - (\ref{eq:22d}) are the result of perturbations to matrices $\bm{O}$ and $\bm{N}$.  These matrices are perturbed due to two effects.  One effect is the set of perturbations due to changes in the solenoidal and quadrupole magnetic fields that appear explicitly in the definitions of $\bm{O}$ and $\bm{N}$.  These are being varied in order to optimize the configuration.  A second effect is due to changes in the self-field terms.  These changes are expressed through the changes in the spatial moments, $\delta \bm{Q}^{(X)}$ in the definitions of Eqs. (\ref{eq:18a}) - (\ref{eq:18c}).  We write the changes to these matrices as a sum of contributions from explicit changes in the focusing configuration and implicit changes in the self-fields,
\begin{subequations}
\label{eq:24}
\begin{align}
\delta \bm{O}^{(X)} \cdot \bm{Q} &= \delta \bm{O}_{Q,B}^{(X)} \cdot \bm{Q} + \bm{M}_Q \cdot \delta \bm{Q}^{(X)}  \, , \label{eq:24a}
\\*[0.2cm]
\delta \bm{O}^{(X)} \cdot \bm{P} &= \delta \bm{O}_{Q,B}^{(X)} \cdot \bm{P} + \bm{M}_P \cdot \delta \bm{Q}^{(X)} \, , \label{eq:24b}
\\*[0.2cm]
\delta \bm{N}^{(X)} &= \delta \bm{N}_{Q,B}^{(X)} + \bm{M}_N \cdot \delta \bm{Q}^{(X)} \, . \label{eq:24c}
\end{align}
\end{subequations}
Here, the variables with subscripts $Q$, $B$ are the contributions to the changes from the explicit dependence of the matrices on the magnetic focusing parameters.  The matrices $\bm{M}_Q,\bm{M}_P,\bm{M}_N$ are the contributions from the self-field terms in the matrices $\bm{M}$, $\bm{N}$, which depend on $\bm{Q}$ as given by Eqs. (\ref{eq:13}), (\ref{eq:14}), and (\ref{eq:18}).  Formulas for these matrices are given in Appendix \ref{A.2}.

Changing each parameter will result in a change to the figure of merit.  This can be thought of as defining a gradient of the figure of merit in parameter space.  Direct evaluation of the change in the figure of merit would require resolving the system of 10 coupled equations $M_p$ times where $M_p$ is the number of parameters that could be varied.  We can avoid this burden by introducing an adjoint system of equations.  Consider a second linear perturbation to the base case, denoted by a superscript ($Y$). These are the adjoint perturbation equations that we will actually solve.
\begin{subequations}
\label{eq:25}
\begin{eqnarray}
\frac{d}{dz}\delta \bm{Q}^{(Y)} &=& \delta \bm{P}^{(Y)} \, , \label{eq:25a}
\\*[0.1cm]
\frac{d}{dz}\delta \bm{P}^{(Y)} &=& \delta \bm{E}^{(Y)} + \bm{O}\cdot \delta \bm{Q}^{(Y)} \, , \label{eq:25b}
\\*[0.1cm]
\frac{d}{dz}\delta \bm{E}^{(Y)} &=& \bm{O}\cdot \delta \bm{P}^{(Y)} + \bm{N}\delta L^{(Y)} + \delta \bm{\dot{E}}^{(Y)} \, , \label{eq:25c}
\\*[0.1cm]
\frac{d}{dz}\delta L &=& -\bm{N}^{\dagger}\cdot\delta \bm{Q}^{(Y)} \label{eq:25d}
\end{eqnarray}
\end{subequations}
In Eqs. (\ref{eq:25a}) - (\ref{eq:25d}) we use the base case (unperturbed) $\bm{O}$ and $\bm{N}$ matrices.  However, we have added a term in Eq. (\ref{eq:25c}), $\delta \bm{\dot{E}}^{(Y)}$, which will be chosen, as described subsequently, so as to achieve a desired cancellation of the self-field contributions to the changes in $\bm{O}$ and $\bm{N}$.

We next form the following combination of true ($X$) and adjoint ($Y$) variables,
\begin{eqnarray} \label{eq:26}
\epsilon \equiv \delta \bm{P}^{(Y)} \cdot \delta \bm{P}^{(X)} - \delta \bm{Q}^{(X)} \cdot \delta \bm{E}^{(Y)} \nonumber \\ - \delta \bm{Q}^{(Y)} \cdot \delta \bm{E}^{(X)} - \delta L^{(Y)} \delta L^{(X)}
\end{eqnarray}
and differentiate it with respect to $z$.  We evaluate the individual terms in the derivative of Eq. (\ref{eq:26}) using the product rule and Eqs. (\ref{eq:22}) and (\ref{eq:25}).  Noting numerous cancellations we arrive at
\begin{align} \label{eq:27}
\frac{d}{dz} \epsilon = \delta \bm{P}^{(Y)} \cdot \delta \bm{O}^{(X)} \cdot \bm{Q} + \delta L^{(Y)}\delta \bm{N}^{\dagger(X)} \cdot \bm{Q} \nonumber \\*[0.1cm] -\delta \bm{Q}^{(X)} \cdot \delta \bm{\dot{E}}^{(Y)} - \delta \bm{Q}^{(Y)} \cdot \delta \bm{O}^{(X)} \cdot \bm{P} - \delta \bm{Q}^{(Y)} \cdot \delta \bm{N}^{(X)}L.
\end{align}
Let us now consider the quantities $\delta \bm{N}^{(X)}$ and $\delta \bm{O}^{(X)}$ appearing in Eq. (\ref{eq:27}). We write them according to the separation defined in Eq. (\ref{eq:22}).  We then have
\begin{align} \label{eq:28}
\frac{d}{dz} \epsilon = \delta \bm{P}^{(Y)} \cdot \bm{M}_Q \cdot \delta \bm{Q}^{(X)} + \delta \bm{P}^{(Y)} \cdot \delta \bm{O}_{Q,B}^{(X)} \cdot \bm{Q} 
\nonumber \\*[0.1cm] -\delta \bm{Q}^{(X)} \cdot \delta \bm{\dot{E}}^{(Y)} + \delta L^{(Y)} \bm{Q} \cdot \bm{M}_N \cdot \delta \bm{Q}^{(X)} 
\nonumber \\*[0.1cm] + \delta L^{(Y)} \bm{Q} \cdot \delta \bm{N}_{Q,B}^{(X)} -\delta \bm{Q}^{(Y)} \cdot \delta \bm{O}_{Q,B}^{(X)} \cdot \bm{P} 
\nonumber \\*[0.1cm] - \delta \bm{Q}^{(Y)} \cdot \bm{M}_P \cdot \delta \bm{Q}^{(X)} - \delta \bm{Q}^{(Y)} \cdot \delta \bm{N}_{Q,B}^{(X)} L 
\nonumber \\*[0.1cm] -\delta \bm{Q}^{(Y)} \cdot \bm{M}_N \cdot \delta \bm{Q}^{(X)} L.
\end{align}
The next step is to pick $\delta \bm{\dot{E}}^{(Y)}$ to cancel all terms proportional to the unknown $\delta \bm{Q}^{(X)}$ in Eq. (\ref{eq:28}),
\begin{align} \label{eq:29}
\delta \bm{\dot{E}}^{(Y)} = \delta \bm{P}^{(Y)} \cdot \bm{M}_Q + \delta L^{(Y)}\bm{Q} \cdot \bm{M}_N
\nonumber \\ -\delta \bm{Q}^{(Y)} \cdot \bm{M}_P - \delta \bm{Q}^{(Y)} \cdot \bm{M}_N L.
\end{align}
This relation is taken to define the quantity $\delta \bm{\dot{E}}^{(Y)}$ that enters the adjoint equations in Eq. (\ref{eq:25c}).  This leaves for the adjoint relation
\begin{align} \label{eq:30}
\frac{d}{dz}\epsilon = \delta \bm{P}^{(Y)} \cdot \delta \bm{O}_{Q,B}^{(X)} \cdot \bm{Q} + \delta L^{(Y)} \bm{Q} \cdot \delta \bm{N}_{Q,B}^{(X)}
\nonumber \\ -\delta \bm{Q}^{(Y)} \cdot \delta \bm{O}_{Q,B}^{(X)} \cdot \bm{P} - \delta \bm{Q}^{(Y)} \cdot \delta \bm{N}_{Q,B}^{(X)} L.
\end{align}
Integrating over $z$ from initial to final point
\begin{widetext}
\begin{eqnarray} \label{eq:31}
\left( \delta \bm{P}^{(Y)} \cdot \delta \bm{P}^{(X)} - \delta \bm{Q}^{(X)} \cdot \delta \bm{E}^{(Y)} - \delta \bm{Q}^{(Y)} \cdot \delta \bm{E}^{(X)} - \delta L^{(Y)} \delta L^{(X)} \right) \Bigg|^{z=z_f}_{z=z_i}
\nonumber \\*[0.2cm] = \int_{z_i}^{z_f} dz \Big\{ \delta \bm{P}^{(Y)} \cdot \delta \bm{O}_{Q,B}^{(X)} \cdot \bm{Q} + \delta L^{(Y)} \bm{Q} \cdot \delta \bm{N}_{Q,B}^{(X)} -\delta \bm{Q}^{(Y)} \cdot \delta \bm{O}_{Q,B}^{(X)} \cdot \bm{P} - \delta \bm{Q}^{(Y)} \cdot \delta \bm{N}_{Q,B}^{(X)} L \Big\}.
\end{eqnarray}
\end{widetext}

We are now in position to describe the utility of the adjoint approach.  Suppose we integrate the base case equations, Eqs. (\ref{eq:12a}) - (\ref{eq:12d}), forward from $z_i$ to $z_f$, and we then integrate the adjoint equations, Eqs. (\ref{eq:25a}) - (\ref{eq:25d}), backward in $z$, starting at $z_f$.  Further, suppose we take the conditions on the adjoint variables at $z_f$ to be
\begin{subequations}
\label{eq:32}
\begin{eqnarray}
\delta \bm{P}^{(Y)}(z_f) &= \frac{\partial F}{\partial \bm{P}} \Big|_{z_f} \, , \label{eq:32a}
\\*[0.1cm]
-\delta \bm{E}^{(Y)}(z_f) &= \frac{\partial F}{\partial \bm{Q}} \Big|_{z_f} \, , \label{eq:32b}
\\*[0.1cm]
-\delta \bm{Q}^{(Y)}(z_f) &= \frac{\partial F}{\partial \bm{E}} \Big|_{z_f} \, , \label{eq:32c}
\\*[0.1cm]
\delta L^{(Y)}(z_f) &= \frac{\partial F}{\partial L} \Big|_{z_f} \, , \label{eq:32d}
\end{eqnarray}
\end{subequations}
where $F$ is the figure of merit.  Then Eqs. (\ref{eq:23}) and (\ref{eq:31}) give for the change in the figure of merit due to perturbation ($X$)

\begin{align} \label{eq:33}
\delta F^{(X)} = \Big( \delta \bm{P}^{(Y)} \cdot \delta \bm{P}^{(X)} - \delta \bm{Q}^{(X)} \cdot \delta \bm{E}^{(Y)} \nonumber \\ - \delta \bm{Q}^{(Y)} \cdot \delta \bm{E}^{(X)} - \delta L^{(Y)} \delta L^{(X)} \Big) \Biggr|_{z=z_i}
\nonumber \\*[0.2cm] + \int_{z_i}^{z_f} dz \Big\{ \delta \bm{P}^{(Y)} \cdot \delta \bm{O}_{Q,B}^{(X)} \cdot \bm{Q} + \delta L^{(Y)} \bm{Q} \cdot \delta \bm{N}_{Q,B}^{(X)} \Big\} 
\nonumber \\*[0.2cm] + \int_{z_i}^{z_f} dz \Big\{ -\delta \bm{Q}^{(Y)} \cdot \delta \bm{O}_{Q,B}^{(X)} \cdot \bm{P} - \delta \bm{Q}^{(Y)} \cdot \delta \bm{N}_{Q,B}^{(X)} L \Big\}.
\end{align}

Equation (\ref{eq:33}) can now be used to evaluate the changes in the figure of merit (FoM) $F$ due to arbitrary, small changes in the initial conditions of the true solution as contained in the first term on the right of Eq. (\ref{eq:33}).  Or it can be used to evaluate changes in the FoM $F$ due to arbitrary, small changes in the strength or profile of the focusing magnetic fields as contained in the integral term on the right of Eq. (\ref{eq:33}).  These changes can be evaluated by a single integral once the adjoint solution is found without having to resolve the coupled system of 10 equations for each possible change in parameters. 

\begin{figure}[t]
\begin{tikzpicture}
\node[] at (0,0) {\includegraphics[width=0.45\textwidth]{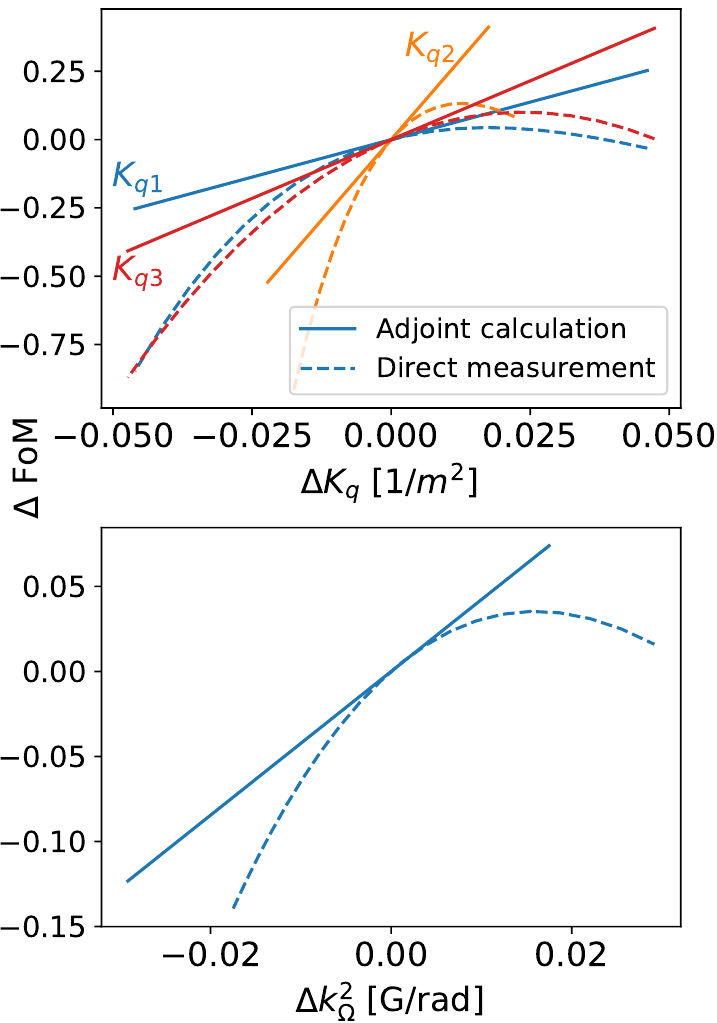}};
\end{tikzpicture}
\caption{Comparison of changes in a figure of merit calculated using the adjoint method (straight lines) and calculated directly (curved lines)}
\label{fig:3}
\end{figure} 

We illustrate the utility of the adjoint method in calculating the gradient of an FoM with respect to variations of parameters in Fig. \ref{fig:3}.  Here we show the dependence on parameters of a particular FoM that will be introduced in the next section.  The dependence on parameter values is calculated two ways: first by directly varying the parameter and second by using Eq. (\ref{eq:33}) to compute the gradient.  The top plot in Fig. \ref{fig:3} shows the variation of the FoM with respect to changes in the strengths of the three quadrupoles, while the bottom plot in Fig. \ref{fig:3} shows the dependence on the solenoidal field.  The solid straight lines have slopes that are predicted by the adjoint method, while the dashed curved lines show the effect of directly varying the parameter and plotting the change in the FoM.  As can be seen, the solid lines are tangent to the dashed lines for vanishing perturbations indicating that the gradient has been calculated correctly. This method of calculating the gradient will be used in the next section to optimize a FTR transformer.

Before using the adjoint method to optimize a configuration we discuss the implications of the two conservation laws following Eq. (\ref{eq:20}). The first conserved quantity we consider is the four-dimensional emittance that is conserved as a consequence of the incompressible, Hamiltonian nature of the particle motion in the phase space, ($x, x', y, y'$). This leads to the constancy of $\epsilon_{4D} = \textrm{det}[\Sigma]$. If we evaluate this quantity upstream for the entering flat beam, assuming all cross moments involving products of $x$ and $y$ are zero, and evaluate it in the solenoid assuming the beam has become cylindrically symmetric (all cross moments are zero and $x$ and $y$ moments are equal), we find:
\begin{align} \label{eq:33.5}
4 \, \textrm{Det}[\Sigma_T] &= 4 \, \langle xx \rangle_u \, \langle yy \rangle_u \, \langle \dot{x}\dot{x} \rangle_u \, \langle \dot{y}\dot{y} \rangle_u \\*[0.2cm]
\nonumber &= (Q_{+s}\,E_{+s} - \frac{1}{2}P_{+s}^2 - \frac{1}{2}L^2)^2 \\*[0.2cm] 
\nonumber &\equiv \epsilon^2_{ts}.
\end{align}
Here the subscript $u$ refers to upstream and subscript $s$ refers to the solenoid. Similarly, evaluating the constant defined in Eq. (\ref{eq:20}), first upstream and then in the solenoid:
\begin{align} \label{eq:33.6}
( \, \langle xx &\rangle_u \, \langle \dot{x}\dot{x} \rangle_u \, + \, \langle yy \rangle_u \, \langle \dot{y}\dot{y} \rangle_u) \\*[0.2cm]
\nonumber &= Q_{+s}\,E_{+s} + \frac{L^2}{2} - \frac{1}{2}P_{+s}^2  \\*[0.2cm]
\nonumber &= \epsilon_{ts} + L^2.
\end{align}
Combining these we can solve for the value of $L$ in the solenoid \cite{kim2003}
\begin{equation} \label{eq:33.7}
L = \pm ( \sqrt{\langle xx \rangle_u \, \langle \dot{x}\dot{x} \rangle_u} - \sqrt{\langle yy \rangle_u \, \langle \dot{y}\dot{y} \rangle_u} ).
\end{equation}
Here the sign of $L$ is determined by the orientation of the quadrupoles that turn the flat beam into a cylindrically symmetric beam. They can be oriented at plus or minus 45 degrees. This follows by noting that changing the orientation of the quadrupoles from plus to minus 45 degrees is equivalent to reversing the sign of the magnetic fields in the quadrupoles. If in addition, one reverses the sign of the solenoidal magnetic field, a configuration that successfully transforms a flat beam to a round beam with one sign of angular momentum, will transform a flat beam to a round beam with the opposite sign of angular momentum.

We have verified that relations Eq. (\ref{eq:33.5}) and (\ref{eq:33.6}) are satisfied by solutions of our moment equations when they apply.  That is, when the beam is cylindrically symmetric in the solenoid. We emphasize that the strength, location, and orientation of the quadrupoles need to be adjusted in the presence of space charge to make the beam cylindrically symmetric in the solenoid.

\section{optimization}

The goal of the optimization is to minimize a given FoM based on the values of the moments at a location ($z=z_f$) inside the solenoid using the adjoint approach. We first choose the following function of the beam moments to be our FoM
\begin{align} \label{eq:34}
F = \frac{1}{2} \Bigg[ |\bm{P}|^2 + k_0^2 (Q_-^2 + Q_x^2) + k_0^{-2}(E_-^2 + E_x^2)
\nonumber \\ k_0^{-2} \Big(E_+ - \frac{1}{2}k_\Omega^2 Q_+ + \Lambda \Big)^2 + (2E_+Q_+ - L^2)^2 \Bigg].
\end{align}
It is a sum of terms quadratic in the moments, each of which should be as small as possible in an FTR transformer that leads to a matched beam in the solenoid.  The quantity $k_0$ is introduced so that each term has the same units.  Here $k_0$ is a scaling parameter approximately equal to the inverse of the lattice length. This makes all the terms roughly comparable in magnitude.  The choice of each term is made as follows.  We would like the beam in the solenoid to be round, $Q_-=Q_x=0$, and we would like all the nonzero moments to be independent of $z$.  This implies for the spatial moments $d\bm{Q}/dz = \bm{P} = 0$.  If we set the components of $d\bm{P}/dz=0$, we find from Eq. (\ref{eq:12b}), $E_-=E_x=0$ is required as well as $E_+ -k_\Omega^2Q_+/2+\Lambda=0$ (radial force balance). Finally, the last term is designed to force the trajectories to be as laminar as possible and the rotation to be rigid in the solenoid.  It can be motivated as follows.  If in the solenoid particles have a radially varying mean rotation rate $\Omega(r)$ in the Larmor frame then
\begin{equation} \label{eq:35}
E_{+0} = \langle \delta x'^2 + \delta y'^2 \rangle + \langle \Omega^2r^2 \rangle
\end{equation}
where $\delta x',\delta y'$ are deviations from the mean rotation. The angular momentum in this case is $L=\langle \Omega r^2 \rangle$. We thus have, by virtue of the Schwarz inequality
\begin{equation} \label{eq:36}
E_+ \geq \langle \Omega^2 r^2 \rangle \geq \langle \Omega r^2 \rangle^2 / \langle r^2 \rangle = L^2 / (2Q_+).
\end{equation}
The second inequality becomes an equality if the rotation is rigid.

The FoM in Eq. (\ref{eq:34}) is a general function in that it meets the requirements for an FTR transformation. In practice an FoM will also contain extra constraint conditions pertinent to the system being optimized. This can include beam size limits based on a physical aperture size, magnet sizes and locations based on available beamline space, etc. The extra constraints can be incorporated within the FoM or as constraints on the input parameters used for the optimization.

\begin{table}[b]
\caption{Entrance and exit conditions for the moments in a thin lens approximation with a symmetric triplet and quadrupoles oriented at 45-degree rotations with respect to the longitudinal direction.}
\begin{ruledtabular}
\begin{tabular}{lr}
$z=0$&
$z=z_f=2d$ \\
\colrule \\
$Q_+=Q_+(0)$ & $Q_+=Q_+(0)$ \\*[0.1cm]
$Q_-=Q_-(0)$ & $Q_-=0$ \\*[0.1cm]
$Q_x=0$ & $Q_x=0$ \\*[0.1cm]
$P_+=0$ & $P_+=0$ \\*[0.1cm]
$P_-=0$ & $P_-=0$ \\*[0.1cm]
$P_x=0$ & $P_x=0$ \\*[0.1cm]
$E_+=Q_+(0) / \big[d^2(2+2\sqrt{2})\big]$ & $E_+=Q_+(0) / \big[d^2(2+2\sqrt{2})\big]$ \\*[0.1cm]
$E_-=Q_-(0) / \big[d^2(2+2\sqrt{2})\big]$ & $E_-=0$ \\*[0.1cm]
$E_x=0$ & $E_x=0$ \\*[0.1cm]
$L=0$ & $L=2Q_-(0) / (d\sqrt{1+\sqrt{2}})$ \\*[0.1cm]
\end{tabular}
\end{ruledtabular}
\label{tab:2}
\end{table}

We start with a design based on the symmetric triplet with parameters given by Eqs. (\ref{eq:1}) and (\ref{eq:2}). It can be shown from analytic solutions of the moment equations in the thin lens approximations that the entrance and exit conditions in Table \ref{tab:2} can be achieved for a symmetric triplet oriented at 45 degrees with respect to the long dimension of the incident flat beam. The strengths of the quadrupoles needed to achieve these parameters are given by
\begin{eqnarray} \label{eq:37}
d\int_{-\infty}^{\infty} 2K_{1,3}(z) s_{1,3} dz = -\sqrt{1+\sqrt{2}} \nonumber \\*[0.2cm]
d\int_{-\infty}^{\infty} 2K_2(z)s_2dz = 2\sqrt{2} / \sqrt{1+\sqrt{2}} \,\, .
\end{eqnarray}
The values given in Eq. (\ref{eq:37}) are in agreement with those expressed in Eq. (\ref{eq:2}).  It is interesting to note that for the 45-degree triplet in the absence of space-charge, the 10 moments break into three independent groups, each with a conserved quantity
\begin{eqnarray} \label{eq:38}
E_-Q_-+L^2/2-P_-^2/2 = J_1 \nonumber \\
(E_+ \pm E_x)(Q_- \pm Q_x) - (P_+ \pm P_x)^2 / 2 = J_\pm \, .
\end{eqnarray}
Also note there are seven quantities which vanish at the exit but only four adjustable parameters, the values of  $E_\pm$ and the strengths of the inner and outer quadrupoles displayed in Eq. (\ref{eq:37}).  The ability to satisfy what seems like three extra conditions then results from the symmetry leading to Eq. (\ref{eq:38}).

\begin{figure*}[t]
\includegraphics[width=\textwidth]{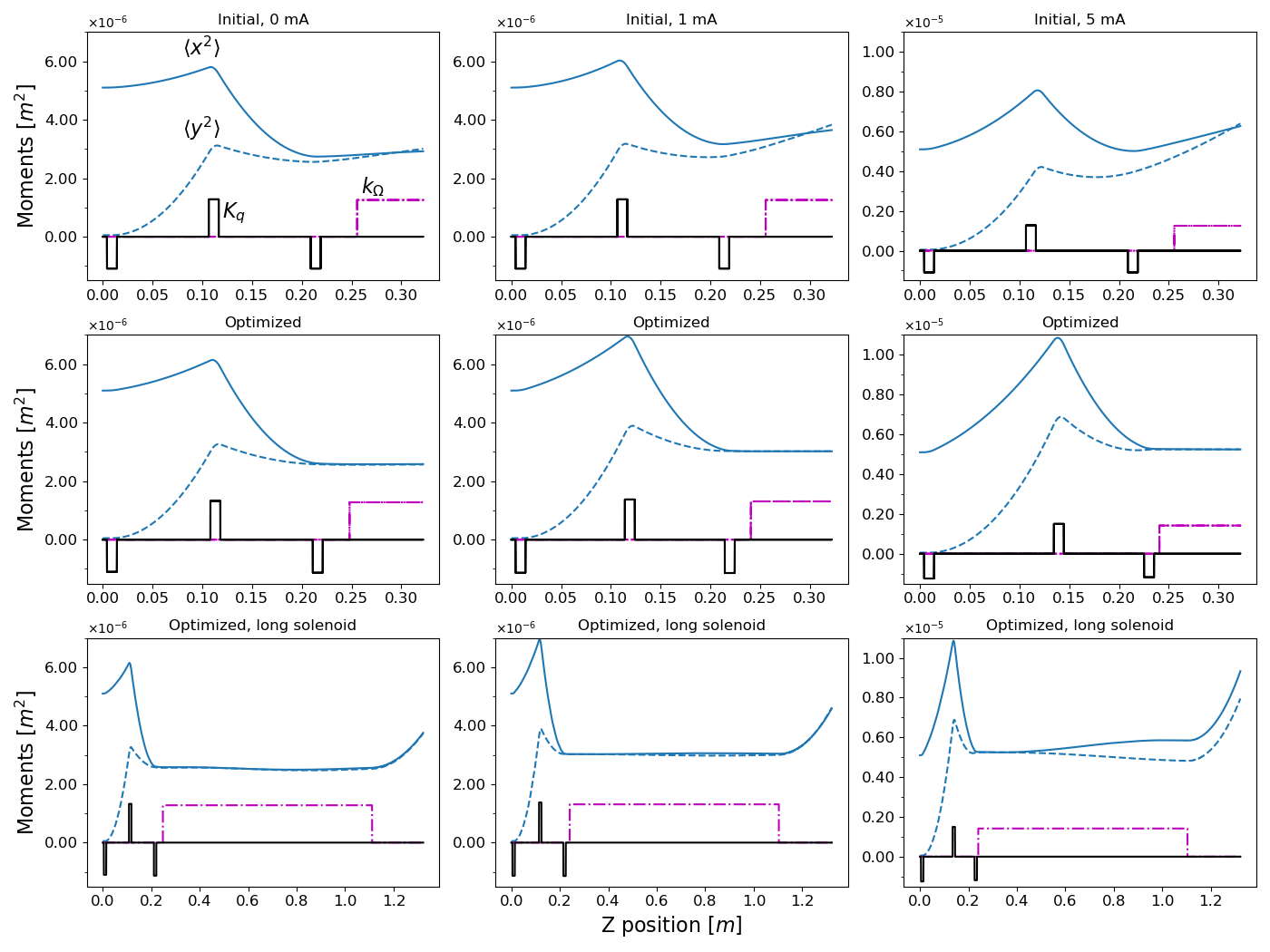}
\caption{RMS beam size, $\langle x^2 \rangle,\langle y^2 \rangle$, plotted as a function of the longitudinal coordinate through a FTR transformer. Top row shows the initial results with no optimization. Middle row shows the optimized results. Bottom row shows the same optimized results as the middle row, but integrates the beam motion through a much longer solenoid (z position). Scaled quadrupole ($K_q$) and solenoid ($k_\Omega$) field profiles are also displayed for reference. Results are for a 5 keV beam.}
\label{fig:4}
\end{figure*}

We now optimize the parameters describing the magnetic fields by using the FoM in Eq. (\ref{eq:34}).  We consider the FoM to be a function of the elements of a list of parameters $\bm{a}$.  The elements of the list include the strengths, locations, and orientations of the quadrupoles as well as the strength and location of the solenoid.  In the case of no self-fields, the strength and location of the solenoid, along with the beam energy, would be enough to optimize the system. With self-fields the initial matching of the beam through the three quadrupoles no longer becomes trivial and requires the quadrupole parameters to be included in the optimization, resulting in a total of 11 parameters. We use a simple steepest descent algorithm in which we calculate the gradient of $F$ in the space of parameters $\bm{a}$, $\nabla_a F(\bm{a})$ and then adjust the values of the parameters by moving from the current set of parameters along the line of steepest descent.  

Generally, we take the parameters $\bm{a}$ to be a set of dimensionless multipliers; each one multiplying the initial value  of its corresponding dimensional quantity. To calculate the gradient we set the conditions at $z=z_f$ for the adjoint variables in Eqs. (\ref{eq:32a}) - (\ref{eq:32d}) according to the current values of the parameters, $\bm{a}_n$. This then allows the solution for the adjoint variables and the integration in Eq. (\ref{eq:33}) to be carried out in order to find the variation in the FoM due to a perturbation in any of the focusing magnets in the system (contained within the $\bm{O}$ and $\bm{N}$ matrices). We then update the parameters according to
\begin{equation} \label{eq:39}
\bm{a}_{n+1} = \bm{a}_n - \gamma \nabla_a F(\bm{a}) \Big|_{a_n} \, .
\end{equation}
The step size $\gamma$ is adjusted iteratively according to a simple algorithm based on successive values of $F(\bm{a}_{n+1}$). Once the values of the parameters $\bm{a}_{n+1}$ are determined the gradient is recomputed and Eq. (\ref{eq:39}) is reapplied. The procedure is repeated until the FoM stops decreasing.

A termination criterion is used in cases where the FoM does not reach a minimum fast enough. Meaning if the improvement in the FoM drops below a certain threshold, then the optimization terminates. A relative tolerance level of around $10^{-7}$ is the cut-off point in the optimizations. At this level the trade-off in computation time for further reduction of FoM is too great. In our case the threshold is acceptable enough for the lattice configurations being investigated. Depending on the lattice accuracy requirements for various experimental setups, the relative tolerance thresholds would need to be adjusted accordingly.


As a first example of the technique, three optimizations are run to optimize a FTR lattice in the presence of self-fields. A 5 keV beam is used with 0 mA (no self-fields), 1 mA, and 5 mA beam currents. For these currents the self-field parameter takes on values $\Lambda=0, \,2.13 \times 10^{-5},$ and $1.06 \times 10^{-4}$.  The importance of self-fields is estimated from Eq. (\ref{eq:19}) to be quantified by the value of  $d^2\Lambda/Q_+(0)=0.0,\, 0.104,$ and $0.518$.   The moment values in Table \ref{tab:3} are used to create a horizontally flat beam with an initial emittance ratio of 100/1.

The lattice consists of a quadrupole triplet with a solenoid at the end. A set of 11 input parameters are tuned for the optimization: the positions and magnet strengths of the three quadrupoles and solenoid along with the rotation angle of each quadrupole. An initial guess at parameters for the 0 mA case is used as the starting values for the optimization. The final optimized results for 0 mA run are then used as initial values for the optimizations with added in self-fields. We can think of the added self-fields as perturbations on the 0 mA solution. Thus, the best starting values for the self-field optimizations are the results obtained from the 0 mA, no self-fields, solution. Results are shown in Fig. \ref{fig:4}.

\begin{table}[t]
\caption{Initial moment values when solving the equations from Eq. \ref{eq:12}}
\begin{ruledtabular}
\begin{tabular}{lr}
Initial Moments ($z=0$)&
Value \\
\colrule \\
$Q_+$ & $2.58 \times 10^{-6}$ \\*[0.1cm]
$Q_-$ & $2.52 \times 10^{-6}$ \\*[0.1cm]
$E_+$ & $5.07 \times 10^{-5}$ \\*[0.1cm]
$E_-$ & $4.97 \times 10^{-5}$ \\*[0.1cm]
\end{tabular}
\end{ruledtabular}
\label{tab:3}
\end{table}


Figure \ref{fig:4} shows that by using the adjoint technique, we are able to find and optimize solutions with increasingly large self-field forces.  Each panel shows the profiles of the $\langle x^2 \rangle$ and $\langle y^2 \rangle$ moments of the beam as well as the location and strength of the quadrupole and solenoidal magnetic fields.  The three columns in Fig. \ref{fig:4} correspond to three different beam currents, 0., 1., and 5. mA.  The top row shows the spatial dependence of the moments with the initial magnet parameters selected based on the thin lens and no space charge approximation.  

The middle row shows the same information after the magnetic fields have been optimized.  The figure of merit is applied at $z = 3.3$m, which is on the right of the panels in the first two rows.  In these optimized cases the beam appears to be round and its moments are constant in the solenoid as demanded by the FoM.  

The bottom row shows the evolution of the moments if propagation of the beam is continued through a longer solenoid, past the point at $z=3.3$m where the FoM is evaluated.  In the 0. And 1. mA cases the beam parameters do not vary after this point and the beam is round indicating the FoM has been minimized to zero.  However, in the 5. mA case it is seen that the beam is not perfectly round in the solenoid.  This is a consequence of our having restricted the solenoid and third quadrupole to not overlap.  When we run the optimizer with no restriction it finds a configuration where the beam is perfectly round in the solenoid.  However, in this case the quadrupole and solenoid overlap.  To avoid this overlap, we do not allow the start location of the solenoidal field to cross to the left of the third quadrupole. 

\begin{figure}[t]
\includegraphics[width=0.45\textwidth]{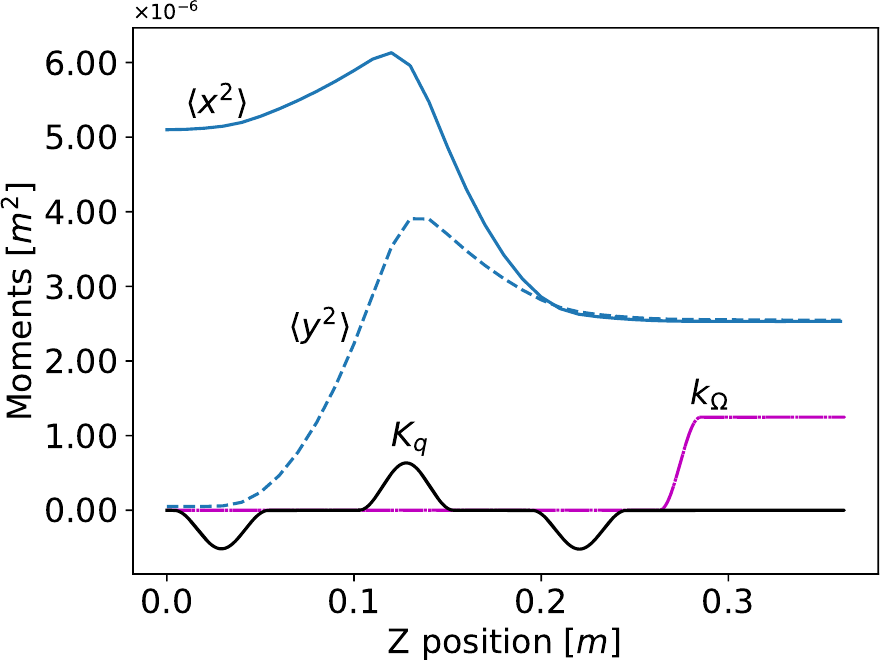}
\caption{RMS beam size vs longitudinal position for an optimized result that uses non-hardedge magnet profiles. A beam with 5 keV energy and zero current is used.}
\label{fig:6}
\end{figure}

While the optimizations in Fig. \ref{fig:4} use hardedge magnet models, the adjoint equations are not limited to such profiles. The magnet strengths, $k_\Omega$ and $K_q$, are continuous variables in the moment equations, and as such, any profile can be used. This may be useful in analyzing certain designs that can not necessarily be represented in matrix form for single particle tracking or would take too long to simulate in PIC tracking codes. Figure \ref{fig:6} shows an optimization result using Gaussian like (actually $\cos^2$) magnet profiles similar to the quadrupole fields available in UMER's magnets \cite{zhang2000}. The solenoid field also has a fringe field edge included in its profile. Various properties of the profiles, such as width, height, etc... can be used as optimization parameters for a given FoM.

\begin{figure}[t]
\includegraphics[width=0.45\textwidth]{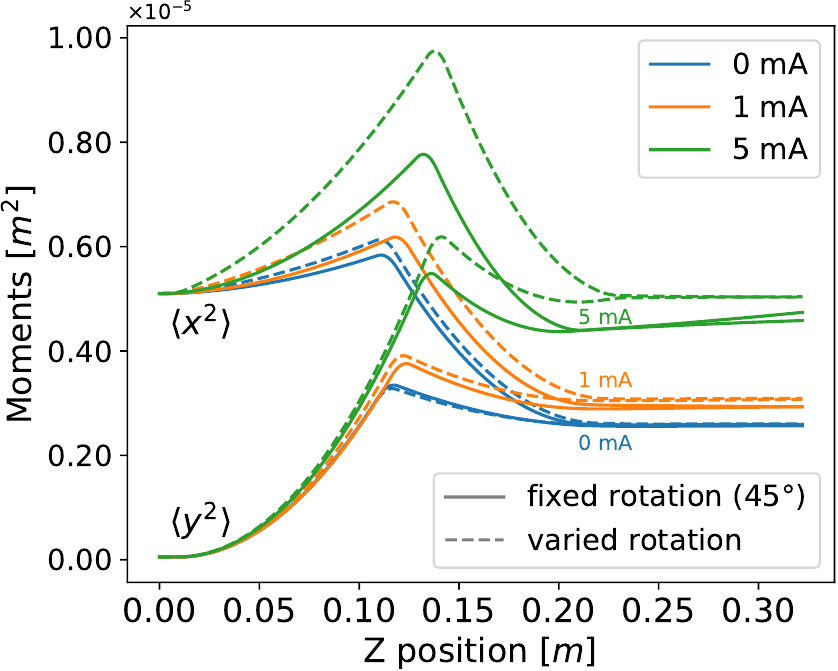}
\caption{FoM vs iterations for optimizations that used varied quadrupole rotation angles (dashed lines) as well as fixed 45-degree rotation angles (solid lines). Optimizations are run for 0 mA, 1 mA, and 5 mA beam currents with 5 keV beam energy.}
\label{fig:8}
\end{figure}

Certain lattice configurations require the quadrupole triplet to maintain a fixed 45-degree transverse rotation angle. While such an FTR solution is possible for a beam with no self-fields, it is not necessarily true when self-fields are added \cite{moroch2021}. Keeping the quadrupole rotation angles fixed, a solution can still be found for beams with little to no space-charge. However, our optimizations show that as the space-charge forces grow, finding an FTR solution becomes progressively more difficult without allowing the quadrupole rotation angles to vary in the optimization, effectively introducing three more tuning parameters that can be used.

Figure \ref{fig:8} demonstrates the effects of fixed vs varied quadrupole rotation angles. A set of optimizations were run using fixed and varied quadrupole rotation angles for three values of space-charge, 0 mA, 1 mA, and 5 mA beams. The results show that as the space-charge force increases, the optimizations made with fixed rotation angles do not have constant moments in the solenoid. Whereas, constant moments are achieved if quadrupole angles are allowed to vary. This shows that in the presence of self-fields the quadrupole triplet matching outside the solenoid is no longer trivial and requires adjusting of rotation angles to properly match the beam into the solenoid.

Another interesting note in Fig \ref{fig:8} is that fixed rotation angle optimizations result in overall smaller beam sizes in the solenoid than varied rotation angle optimizations. All the optimizations were run with the same initial beam conditions (beam size, emittance, etc...). We also observed that tuning the initial conditions did not significantly improve the optimized results in the cases of fixed rotation angles. Our results, through a set of numerical optimizations, show that, in agreement with \cite{moroch2021}, with increasing space-charge forces the quadrupole rotations angles need to be tuned in order to meet FTR transformation conditions.

All optimizations up to this point have used the same initial beam conditions displayed in Table \ref{tab:2}. These initial conditions are needed in order to achieve a successful FTR transformation. In particular, the values of $E_+$ and $E_-$, which depend on $Q_+$ and $Q_-$. However, some level of deviation from these initial conditions is expected in experiments. It might be difficult to exactly produce a beam with the required initial conditions. Figure \ref{fig:9} shows that it is possible to correct for mismatched parameters using the optimizer.  The three panels correspond to three different current values.  Each panel displays three pairs of profiles for the spatial second moments.  The profiles marked with the circles are the optimized profiles from the middle row of Fig. \ref{fig:4}.  These were constructed with initial values for $E_+$ and $E_-$ given in table II.  We then changed the $E_{+,-}$ moment values by 50\%.  This resulted in the pairs of red curves which indicate the beam is not round in the solenoid.  The optimizer was then run with the perturbed moments and the beam was brought back to round as shown by the blue curves.

The FoM in Eq. \ref{eq:34} is constructed as a sum of terms each of which can be made to vanish. As mentioned the terms are selected to make the moments independent of axial distance in the solenoid.  The second to last term enforces radial force balance so that $P_+$ is constant.  The last term enforces rigid rotation.  This term has had little effect on the optimizations shown so far, so we omit it from further considerations.  If the beam is in radial force balance it must acquire a mean rotation to balance the radial self-fields.  This would degrade the cooling properties for a co-propagating beam. To show the robustness of the optimization technique and explore a trade-off between cooling potential and uniformity of moments in the solenoid, we modify the FoM by adding a new fifth term that reflects the transverse energy of the beam particles in the lab frame.  The new FoM is:

\begin{figure*}[!htbp]
\includegraphics[width=\textwidth]{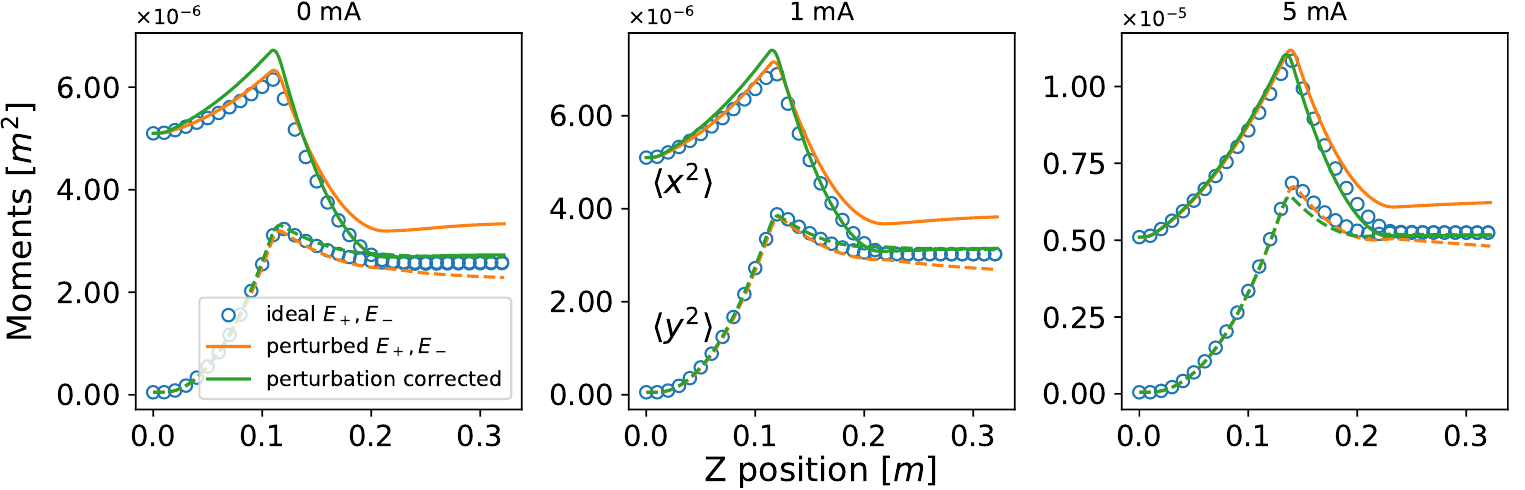}
\caption{RMS beam size is plotted for optimized results with ideal initial conditions (blue circles). Initial conditions are purposely perturbed to break the optimized results (orange line). Magnet parameters are re-optimized to correct for the perturbed initial conditions (green line).}
\label{fig:9}
\end{figure*}

\begin{figure*}[t]
\includegraphics[width=1.0\textwidth]{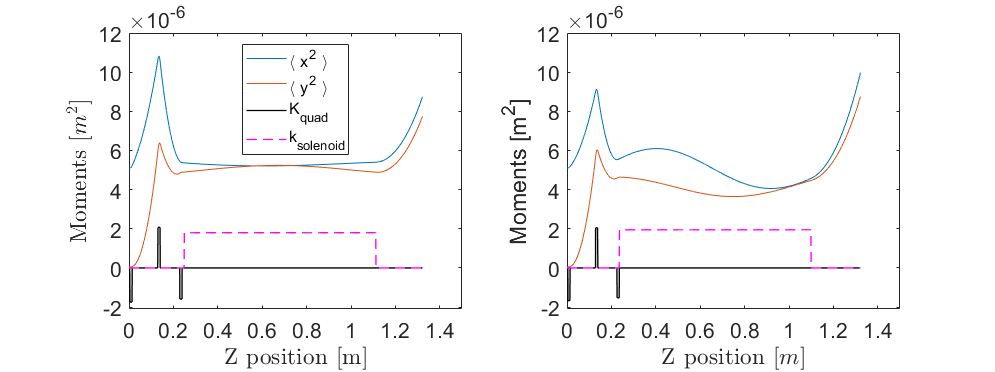}
\caption{RMS beam size is plotted for two different optimization using the FoM from Eq. \ref{eq:44}. The left side uses weight values of $\epsilon_1,\epsilon_2 = 1,0$ while the right side uses $\epsilon_1,\epsilon_2 = 0,1$. A 5 mA beam current and the same set of initial conditions is used for both optimizations. The optimizations minimized the FoM from Eq. \ref{eq:44} at roughly the center of the solenoid ($Z = 0.722$ m).}
\label{fig:10}
\end{figure*}

\begin{align} \label{eq:44}
F_\textrm{total} = F_1 + F_2 + F_3 + \epsilon_1 F_4 + \epsilon_2 F_5
\end{align}
where 
\begin{subequations}
\label{eq:a1}
\begin{eqnarray}
F_1 &=& \frac{1}{2} |\bm{P}|^2  \, , \label{eq:45a}
\\*[0.1cm]
F_2 &=& \frac{1}{2} k_0^2 (Q_-^2 + Q_x^2)  \, , \label{eq:45b}
\\*[0.1cm]
F_3 &=& \frac{1}{2} k_0^{-2}(E_-^2 + E_x^2)  \, , \label{eq:45c}
\\*[0.1cm]
F_4 &=&  \frac{1}{2} k_0^{-2} \Big(E_+ - \frac{1}{2}k_\Omega^2 Q_+ + \Lambda \Big)^2 \, , \label{eq:45d}
\\*[0.1cm]
F_5 &=& \frac{1}{2} k_0^{-2} E_{+\textrm{lab}}^2 \label{eq:45e}
\end{eqnarray}
\end{subequations}
and $E_{+\textrm{lab}}$ is the $E_+$ moment from Eq. \ref{eq:11} represented in the lab frame (vs the Larmor frame) as:

\begin{equation} \label{eq:45}
E_{+\textrm{lab}} = (E_+ + \frac{1}{2}k^2_\Omega Q_+ - k_\Omega L).
\end{equation}

The terms $F_1-F_3$ measure the degree of axial uniformity and the degree of cylindrical symmetry of the beam in the solenoid. The fourth term, $F_4$, in the FoM still represents radial force balance. These terms can, in principle, be minimized to zero. The fifth term, $F_5$, which is new, is positive definite and cannot be minimized to zero; it represents the beam's transverse kinetic energy in the lab frame and should be minimized if the considered application is co-propagating hadron beam cooling. In principle, one would like to minimize the integral of the transverse energy over the length of the solenoid rather than in just one plane as done here. Further, one would also want to account for the radial size and distribution of both the hadron beam and the cooling beam. However, this is beyond the scope of the present paper, and we will simply minimize the transverse energy in a single plane. Further, a pair of weighting parameters labeled $\epsilon_1 , \epsilon_2$ is included in the FoM in order to trade-off between minimizing the beam's radial oscillations, as measured by $F_4$, and transverse energy, as measured by $F_5$, within the solenoid.

\begin{table}[b]
\caption{Optimization results for minimizing components $F_4$ and $F_5$ from the FoM in Eq. \ref{eq:44}.}
\begin{ruledtabular}
\begin{tabular}{ccc}
FoM & $\epsilon_1,\epsilon_2 = 1,0$ & $\epsilon_1,\epsilon_2 = 0,1$ \\
\colrule \\*[-0.1cm]
$F_1$ & $0.0005 \times 10^{-10}$ & $0.0812 \times 10^{-10}$ \\*[0.1cm]
$F_2$ & $0.0000 \times 10^{-10}$ & $0.1212 \times 10^{-10}$ \\*[0.1cm]
$F_3$ & $0.0026 \times 10^{-10}$ & $0.0172 \times 10^{-10}$ \\*[0.1cm]
$F_4$ & $0.0003 \times 10^{-10}$ & $0.0267 \times 10^{-10}$ \\*[0.1cm] 
$F_5$ & $0.3100 \times 10^{-10}$ & $0.1986 \times 10^{-10}$ \\*[0.1cm]
\end{tabular}
\end{ruledtabular}
\label{tab:4}
\end{table}

Using the new FoM in Eq. \ref{eq:44} along with a 5 mA high space-charge beam, an optimization is performed using weighting parameters $\epsilon_1,\epsilon_2$ to prioritize either minimizing the beam's radial oscillations or its transverse energy within the solenoid. The results are shown in Fig. \ref{fig:10}. On the left side of Fig. \ref{fig:10} the weight parameters are set to $\epsilon_1,\epsilon_2 = 1,0$ which forces the optimization to focus on minimizing the beam's radial oscillations, creating a relatively flat, oscillation free, beam within the solenoid. On the right side of Fig \ref{fig:10} the weight parameters are set to $\epsilon_1,\epsilon_2 = 0,1$ which directs the optimization to instead minimize the beam's transverse energy. Table \ref{tab:4} shows the calculated FoM component values post optimization with the indicated weight parameters. That is, the two optimizations were performed with weight parameters set to $\epsilon_1,\epsilon_2 = (1,0)$ and $(0,1)$. Using the optimized magnet parameters, the weight parameters were reset to $\epsilon_1,\epsilon_2 = (1,1)$ and each FoM component was calculated; these are the values shown in Table \ref{tab:4}. The table shows that the transverse energy, $F_5$, is lower (by a third) for the optimization of $\epsilon_1,\epsilon_2 = (0,1)$ than for the $\epsilon_1,\epsilon_2 = (1,0)$ optimization as expected. The beam's radial oscillations, $F_4$, along with the other FoM components, are larger in the $\epsilon_1,\epsilon_2 = (0,1)$ optimization, again as expected. For the optimization case of $\epsilon_1,\epsilon_2 = (1,0)$, the first four components of the FoM are very small, but the transverse energy term, $F_5$, is bigger than the first four in this case.

Using the modified FoM in Eq. \ref{eq:44}, the optimization results in Fig. \ref{fig:10} and Table \ref{tab:4} demonstrate the ability to trade-off minimization of different properties of the beam within the solenoid. Depending on a given experimental tolerance, a pair of weight parameters can be used to create a balance between the beam's radial oscillations and transverse energy within the solenoid. Such a method can be applied to any set of terms in a given FoM.

\section{conclusion}

In conclusion, this paper presents the adjoint approach to calculating the gradient with respect to magnet parameters of figures of merit characterizing accelerator lattices. The gradient is then used in a steepest descent algorithm to optimize a Flat-to-Round lattice transformer in the presence of space-charge forces. The optimization is developed for a model of the system consisting of a set of continuous differential equations for the 10 moments describing the 4D phase space in the paraxial limit. The optimization is achieved by tuning a set of quadrupole and solenoid magnet parameters within the lattice.

The adjoint approach and gradient descent algorithm is able to successfully optimize various transformer systems. The optimization can account for space-charge forces, continuous variation of magnetic field profiles, and for mismatches in beam parameters on entry to the transformer. In using the adjoint approach to the problem, optimizations are performed at significantly reduced computational costs. The system of moment equations can be solved twice and the gradients with respect to all parameters determined as integrals of the adjoint solution. Direct determination of the gradient would require as many simulations as there are parameters.

While the results shown here are based on a reduced model of the charged particle dynamics, namely a solution of the moment equations assuming linear restoring forces, the adjoint method can also be applied to a particle description (see \cite{antonsen2019} and \cite{antonsen2019_2}).  In further work we plan to generalize the optimization procedure to the particle description of the beam dynamics, and to optimization of circular lattices.  A future challenge is to include dissipative effects such as synchrotron radiation that are important for high energy accelerators. 

\section{acknowledgements}

This work was supported by DOE-HEP award No. DE-SC0010301 and DE-SC0022009.

\begin{appendix}

\section{COUPLED MOMENT EQUATIONS} \label{A.1}

In this appendix we outline the steps leading to Eqs. (\ref{eq:12})-(\ref{eq:14}).  We start with the definitions of the moments in Eqs. (\ref{eq:8}) - (\ref{eq:11}).  Equation (\ref{eq:12a}), the derivative of $\bm{Q}$ with respect to distance $z$, follows directly from the definition of the components of $\bm{P}$ in Eq.  (\ref{eq:9}).  Differentiating the vector $\bm{P}$ and angular momentum $L$ with respect to $z$ gives,
\begin{subequations}
\label{eq:a1}
\begin{eqnarray}
P_+' &=& E_+ + \langle xx'' + yy'' \rangle \, , \label{eq:a1a}
\\*[0.1cm]
P_-' &=& E_- + \langle xx'' - yy'' \rangle  \, , \label{eq:a1b}
\\*[0.1cm]
P_x' &=& E_x + \langle xy'' + yx'' \rangle  \, , \label{eq:a1c}
\\*[0.1cm]
L' &=& \langle xy'' - yx'' \rangle  \label{eq:a1d}
\end{eqnarray}
\end{subequations}
where the moments $\bm{E}$ are defined in Eq. (\ref{eq:11}).  Differentiating the moments $\bm{E}$ with respect to axial distance gives
\begin{subequations}
\label{eq:a2}
\begin{eqnarray}
E_+'= 2\langle x'x'' + y'y'' \rangle \, , \label{eq:a2a}
\\*[0.1cm]
E_-'= 2\langle x'x'' - y'y'' \rangle  \, , \label{eq:a2b}
\\*[0.1cm]
E_x'= 2\langle x'y'' + y'x'' \rangle  \, , \label{eq:a2c}
\end{eqnarray}
\end{subequations}
Evaluation of each of the averages on the right sides of Eqs. (\ref{eq:a1}) and (\ref{eq:a2}) requires inserting expressions for the second derivatives of $x$ and $y$ in the Larmor Frame.  For this we use Eq. (\ref{eq:8}).

The first contribution to the averages in Eqs. (\ref{eq:a1}) and (\ref{eq:a2}) comes from the
solenoidal field contribution to Eq. (\ref{eq:8}),
\begin{equation} \label{eq:a3}
\begin{pmatrix}[1.5] x_B''\\ y_B'' \end{pmatrix} = -\Bigg(\frac{k_\Omega}{2}\Bigg)^2 \begin{pmatrix}[1.5] x \\ y \end{pmatrix}
\end{equation}
This when inserted in Eqs. (\ref{eq:a1}) and (\ref{eq:a2}) gives rise to the first term in the expression for the matrix $\bm{O}$ in Eq. (\ref{eq:13}). 

The second contribution to the averages in Eqs. (\ref{eq:a1}) and (\ref{eq:a2}) comes from the quadrupole fields. To evaluate those terms we write the second derivatives in the Larmor frame in the following way
\begin{align} \label{eq:a4}
\begin{pmatrix}[1.5] x_Q''\\ y_Q'' \end{pmatrix} &= \nonumber\\ \bm{R}^{-1}(\phi) & \sum_{\textrm{Quads}}K_Q \begin{pmatrix}[1.5] \cos{2\psi} & \sin{2\psi} \\ \sin{2\psi} & -\cos{2\psi} \end{pmatrix} \bm{R}(\phi) \begin{pmatrix}[1.5] x \\ y \end{pmatrix}
\end{align}
Here the Larmor variables, $x$ and $y$, are first transformed back to the lab frame, where Eq. (\ref{eq:16}) describes the quadrupole fields.  The transverse Lorentz force is computed and the lab frame, and the resulting acceleration is transformed to the Larmor frame.  Multiplying the three matrices in Eq. (\ref{eq:a4}) gives
\begin{align} \label{eq:a5}
\bm{R}^{-1}(\phi) \sum_{\textrm{Quads}}K_Q \begin{pmatrix}[1.5] \cos{2\psi} & \sin{2\psi} \\ \sin{2\psi} & -\cos{2\psi} \end{pmatrix} \bm{R}(\phi)
\nonumber \\ 
= -\sum_{\textrm{Quads}}K_Q \begin{bmatrix}[1.5] -\cos{(2\phi-2\psi)} & \sin{(2\phi-2\psi)} \\ \sin{(2\phi-2\psi)} & \cos{(2\phi-2\psi)} \end{bmatrix}
\end{align}
Inserting Eq. (\ref{eq:a5}) into Eq. (\ref{eq:a4}) and then inserting the resulting second derivatives in the expressions in Eq. (\ref{eq:a1}) and (\ref{eq:a2}) for the rates of change of the moments gives rise to the second contribution to the matrix O in Eq. (\ref{eq:13}) and the first contribution to the vector N in Eq. (\ref{eq:14}).

The evaluation of the self-field force is done in what we label the beam frame.  In the beam frame the beam is assumed to have an elliptical shape with major and minor semi-axes radii $a$ and $b$, aligned with the $x_2$ and $y_2$ directions respectively.  The $x_2$ direction makes an angle $\alpha$ with respect to the x-axis in the Larmor frame.  The transformation from beam frame to Larmor frame coordinates is thus given by
\begin{equation} \label{eq:a6}
\begin{pmatrix}[1.5] x \\ y \end{pmatrix} = \bm{R}(\phi) \begin{pmatrix}[1.5] x_2 \\ y_2 \end{pmatrix}
\end{equation}
The beam is assumed to have uniform density within an elliptical cross section in the beam frame.  Thus, the beam frame moments satisfy,
\begin{equation} \label{eq:a7}
\langle x_2^2 \rangle = a^2 / 4 \, , \, \langle y_2^2 \rangle = b^2/4 \, , \, \langle x_2y_2 \rangle = 0
\end{equation}
This implies through application of Eq. (\ref{eq:a6}) for the Larmor frame moments,
\begin{align} \label{eq:a8}
\langle x^2 \rangle &= \frac{1}{8} \Big[ (a^2 + b^2) + \cos{(2\alpha)}(a^2-b^2)\Big] \nonumber \\*[0.1cm]
\langle y^2 \rangle &= \frac{1}{8} \Big[ (a^2 + b^2) - \cos{(2\alpha)}(a^2-b^2)\Big] \nonumber \\*[0.1cm]
\langle xy \rangle &= \frac{1}{8} (a^2-b^2) \sin{(2\alpha)}
\end{align}
The Larmor frame values for the elements of $\bm{Q}$ are then related to the radii $a$ and $b$ and the angle $\alpha$ by
\begin{align} \label{eq:a9}
Q_+ &= \frac{1}{2} \langle x^2 + y^2 \rangle = \frac{1}{8}(a^2+b^2) \nonumber \\*[0.1cm]
Q_- &= \frac{1}{2} \langle x^2 - y^2 \rangle = \frac{\cos{(2\alpha)}}{8} (a^2-b^2) \nonumber \\*[0.1cm]
Q_x &= \langle xy \rangle = \frac{\sin{(2\alpha)}}{8} (a^2-b^2)
\end{align}

\section{SELF-FIELDS AND TRANSFORMATIONS} \label{A.2}

The effective electric potential due to a beam with elliptical cross section far from conducting boundaries is given by
\begin{widetext}
\begin{align} \label{eq:b1}
\Phi = - \frac{2\Lambda}{(a+b)}\left[ \frac{x_2^2}{a} + \frac{y_2^2}{b} \right]
= -\frac{2\Lambda}{ab} \left[ \frac{1}{2} (x^2 + y^2) + \frac{b-a}{b+a} \left( \frac{\cos{(2\alpha)}}{2}(x^2-y^2) + \sin{(2\alpha)}xy \right) \right]
\end{align}
\end{widetext}
where the first expression is in the beam frame and the second expression has been transformed to the Larmor frame.  The current strength parameter $\Lambda$ is defined in Eq. (\ref{eq:19}).  The relation between the effective potential and the space-charge potential $\phi_{\text{sc}}$ is $\Phi=q\phi_{\textrm{sc}}/(m\nu_z^2\gamma^3)$. One of the three powers of $\gamma$ in the denominator of the effective potential accounts for the effective relativistic mass increase of the beam particles, and the other two powers accounts for the self magnetic field.  The accelerations due to self-fields then follow from
\begin{equation} \label{eq:b2}
\begin{pmatrix}[1.5] x_\Lambda'' \\ y_\Lambda'' \end{pmatrix} = - \begin{pmatrix}[1.5] \partial/\partial x \\ \partial / \partial y \end{pmatrix} \Phi(x,y)
\end{equation}
When Eq. (\ref{eq:b2}) is inserted in Eqs. (\ref{eq:a1}) and (\ref{eq:a2}), and the parameters $a$, $b$, and $\alpha$ are expressed in terms of the moments $\bm{Q}$ using Eq. (\ref{eq:a9}), the result is the self-field contributions to the matrix $\bm{O}$ and vector $\bm{N}$ appearing in Eqs. (\ref{eq:13}) and (\ref{eq:14}) along with definitions in Eq. (\ref{eq:18}).

Equations (\ref{eq:12a})-(\ref{eq:12d}) describe the evolution of the moments in the Larmor frame.  If the values of the moments in the lab frame are desired, these can be recovered by the following transformations,
\begin{subequations}
\label{eq:b3}
\begin{align} 
\bm{Q}_{\textrm{lab}} &= \bm{R}_1 \cdot \bm{Q} \label{eq:b3a} \\*[0.1cm]
\bm{P}_{\textrm{lab}} &= \bm{R}_1 \cdot \bm{P} + 2\phi' \bm{R}_2 \cdot \bm{Q} \label{eq:b3b}\\*[0.1cm]
\bm{E}_{\textrm{lab}} &= \bm{R}_1 \cdot \bm{E} + 2\phi'\bm{R}_2 \cdot \bm{P} + 2\phi'^2\bm{R}_3 \cdot \bm{Q} + 2\phi'\bm{1}_+ L \label{eq:b3c}\\*[0.1cm]
L_{\textrm{lab}} &= L + 2\phi'Q_+ \label{eq:b3d}
\end{align}
\end{subequations}
Here the following matrices are defined
\begin{equation} \tag{B4a}
\bm{R}_1 = \begin{bmatrix}[1.5] 1&0&0\\0&\cos{(2\phi)}&-\sin{(2\phi)}\\0&\sin{(2\phi)}&\cos{(2\phi)} \end{bmatrix} \label{eq:b4a} \\*[0.1cm]
\end{equation}
\begin{equation} \tag{B4b}
\bm{R}_2 = \begin{bmatrix}[1.5] 0&0&0\\0&-\sin{(2\phi)}&-\cos{(2\phi)}\\0&\cos{(2\phi)}&-\sin{(2\phi)} \end{bmatrix} \label{eq:b4b} \\*[0.1cm]
\end{equation}
\begin{equation} \tag{B4c}
\bm{R}_3 = \begin{bmatrix}[1.5] 1&0&0\\0&-\cos{(2\phi)}&\sin{(2\phi)}\\0&-\sin{(2\phi)}&-\cos{(2\phi)} \end{bmatrix} \label{eq:b4c} \\*[0.1cm]
\end{equation}
\begin{equation} \tag{B4d}
\bm{1}_+ = \begin{pmatrix}[1.5] 1\\0\\0 \end{pmatrix} \label{eq:b4d}
\end{equation}

The adjoint equation (\ref{eq:25c}) has an added term $\delta\bm{\dot{E}}^{(Y)}$, which is added to cancel out the terms in Eq. (\ref{eq:27}) that come from the dependence of the matrix $\bm{O}$ and vector $\bm{N}$ on the changes in the self-fields.  These changes are proportional to changes in the moments $\delta\bm{Q}^{(X)}$.  To represent these changes in the $\bm{O}$ matrix we first separate the changes due to the self-fields ($\Lambda$) from the changes in the magnetic focusing parameters ($Q,B$), $\delta\bm{O}^{(X)}=\delta\bm{O}_{Q,B}^{(X)}+\delta\bm{O}_\Lambda^{(X)}$ .  We then construct a matrices $\bm{M}_P$ and $\bm{M}_Q$ that satisfy
\begin{subequations}
\begin{align}
\delta\bm{O}_\Lambda^{(X)} \cdot \bm{P} = \bm{M}_P \cdot \delta \bm{Q}^{(X)} \tag{B5a} \label{eq:b5a} \\*[0.1cm]
\delta\bm{O}_\Lambda^{(X)} \cdot \bm{Q} = \bm{M}_Q \cdot \delta \bm{Q}^{(X)} \tag{B5b} \label{eq:b5b} 
\end{align}
\end{subequations}
The vector $N$ is treated similarly.

We focus now on $\bm{M}_P$. We start by forming the vector
\begin{equation} \tag{B6} \label{eq:b6}
\bm{R}_P = \bm{O}_\Lambda \cdot \bm{P} = \frac{\Lambda}{Q_\Delta} \begin{pmatrix}[1.5] P_+-\frac{Q_-}{Q_\Delta + Q_+}P_--\frac{Q_x}{Q_\Delta+Q_+}P_x \\ -\frac{Q_-}{Q_\Delta+Q_+}P_+ + P_- \\ -\frac{Q_x}{Q_\Delta+Q_+}P_+ + P_x \end{pmatrix}
\end{equation}
The elements of the matrix $\bm{M}_P$ are generated by differentiating each element of vector $\bm{R}_P$ with respect to each element of vector $\bm{Q}$,
\begin{equation} \tag{B7} \label{eq:b7}
\bm{M}_P = \left[ \frac{\partial \bm{R}_P}{\partial \bm{Q}} \right]^T
\end{equation}
To keep track of terms we decompose $\bm{M}_P$ according to
\begin{equation} \tag{B8} \label{eq:b8}
\bm{M}_P = \bm{V}_1\bm{U}_1^T + \bm{V}_2\bm{U}_2^T + \bm{V}_3\bm{U}_3^T + \bm{V}_4\bm{U}_4^T
\end{equation}
where
\begin{equation} \tag{B9a} \label{eq:b9a}
\bm{V}_1 = -\frac{\Lambda}{Q_\Delta^2} \begin{pmatrix}[1.5] P_+-\frac{Q_-}{Q_\Delta + Q_+}P_--\frac{Q_x}{Q_\Delta+Q_+}P_x \\ -\frac{Q_-}{Q_\Delta+Q_+}P_+ + P_- \\ -\frac{Q_x}{Q_\Delta+Q_+}P_+ + P_x \end{pmatrix}
\end{equation}
\begin{equation} \tag{B9b} \label{eq:b9b}
\bm{U}_1^T = \begin{pmatrix} \frac{\partial Q_\Delta}{\partial Q_+} & \frac{\partial Q_\Delta}{\partial Q_-} & \frac{\partial Q_\Delta}{\partial Q_x} \end{pmatrix} = \frac{1}{Q_\Delta} \begin{pmatrix} Q_+ & -Q_- & -Q_x \end{pmatrix}
\end{equation}
\begin{equation} \tag{B9c} \label{eq:b9c}
\bm{V}_2 = \frac{\Lambda}{Q_\Delta(Q_\Delta+Q_+)^2} \begin{pmatrix}[1.5] Q_-P_- + Q_xP_x \\ Q_-P_+ \\ Q_xP_+ \end{pmatrix}
\end{equation}
\begin{align} \nonumber \label{eq:b9d}
\bm{U}_2^T &= \begin{pmatrix} \frac{\partial (Q_\Delta+Q_+)}{\partial Q_+} & \frac{\partial (Q_\Delta+Q_+)}{\partial Q_-} & \frac{\partial (Q_\Delta+Q_+)}{\partial Q_x} \end{pmatrix} \\ \tag{B9d} &= \bm{U}_1^T +\begin{pmatrix} 1&0&0 \end{pmatrix}
\end{align}
\begin{align} \nonumber\label{eq:b9e}
\bm{V}_3 = - \frac{\Lambda}{Q_\Delta(Q_\Delta+Q_+)} \begin{pmatrix}[1.5] P_- \\ P_+ \\ 0 \end{pmatrix} ,
\\ \tag{B9e}  \bm{V}_4 = - \frac{\Lambda}{Q_\Delta(Q_\Delta+Q_+)} \begin{pmatrix}[1.5] P_x \\ 0 \\ P_+ \end{pmatrix}
\end{align}
\begin{equation} \tag{B9f} \label{eq:b9f}
\bm{U}_3^T = \begin{pmatrix} 0&1&0 \end{pmatrix} , \, \bm{U}_4^T = \begin{pmatrix} 0&0&1 \end{pmatrix}
\end{equation}
The matrices $\bm{M}_Q$ and $\bm{M}_N$ are constructed in the same way.

\section{SCALING PROPERTIES} \label{A.3}
Finally, we discuss the scaling properties of our system of moment equations.  Let us suppose that we have found a solution to the moment equations in the form of functions
\begin{equation} \label{eq:c1}
\bm{Q}_0(z_0),\bm{P}_0(z_0),\bm{E}_0(z_0),\bm{O}_0(z_0),\bm{N}_0(z_0),L_0(z_0)
\end{equation}
It can then be verified that a scaled solution that is also satisfying the moment equations is
\begin{align} \label{eq:c2}
\bm{Q}(z) = \lambda \bm{Q}_0(\epsilon z) \nonumber \\*[0.1cm]
\bm{P}(z) = \lambda \epsilon \bm{P}_0(\epsilon z) \nonumber \\*[0.1cm]
\bm{E}(z) = \lambda \epsilon^2 \bm{E}_0(\epsilon z) \nonumber \\*[0.1cm]
\bm{O}(z) = \epsilon^2 \bm{O}_0(\epsilon z) \nonumber \\*[0.1cm]
\bm{N}(z) = \epsilon^2 \bm{N}_0(\epsilon z) \nonumber \\*[0.1cm]
L(z) = \lambda \epsilon L_0(\epsilon z) 
\end{align}
Here $\epsilon,\lambda$ are arbitrary constants allowing for separate scaling of the axial dependence and size of the beam.  If $\epsilon>1$ the scaled solution is shorter in spatial length than the original solution, and if $\epsilon<1$ the scaled solution is longer than the original.  If $\lambda>1$ the scaled beam is larger.  Of course this is subject to the requirement that the transverse forces remain linear in displacements.

Looking at the expressions for the matrices $\bm{O},\bm{N}$ in Eq. (\ref{eq:c2}) , there are a number of conditions that must be satisfied to give the required $\epsilon^2$ scaling.  The solenoidal field contribution requires,
\begin{equation} \label{eq:c3}
k_\Omega(z) = \epsilon \,k_{\Omega 0}(\epsilon z)
\end{equation}
This means a shorter solution requires a stronger solenoidal field.  The scaled solution gives for the phase
\begin{equation} \label{eq:c4}
\phi(z) = \phi_0 (\epsilon z)
\end{equation}
Thus, the values of the phase in the locations of the quadrupoles are preserved under the scaling.  For the quadrupoles, strict application of the scaling gives
\begin{equation} \label{eq:c5}
K_q(z) = \epsilon^2 K_{q0}(\epsilon z)
\end{equation}
However, in the thin lens approximation only the integrated value of the quadrupole field matters
\begin{equation} \label{eq:c6}
K = \int dz \, K_q(z) = \epsilon \int \epsilon \, dz \, K_{q0}(\epsilon z) = \epsilon K_0
\end{equation}
In this approximation the strength of the quadrupole field also scales inversely with length. 

The self-field contribution to the $\bm{O}$ and $\bm{N}$ matrices can also be preserved.  We note that these contributions scale as
\begin{equation} \label{eq:c7}
O,N \propto \Lambda / Q
\end{equation}
So to preserve the solution we require
\begin{equation} \label{eq:c8}
O \propto \Lambda / Q = \Lambda / (\epsilon \lambda Q_0) = \epsilon^2 O_0 \propto \epsilon^2 \Lambda_0 / Q_0
\end{equation}
As a result, the current parameter scales as
\begin{equation} \label{eq:c9}
\Lambda = \lambda \epsilon^3 \Lambda_0
\end{equation}
Thus, scaling length or amplitude requires changing beam current to maintain space-charge influence.

\end{appendix}



\bibliography{main}

\end{document}